# Chapter 06

# How Does Incubation Affect Laser Material Processing?

**Matthias Lenzner[1,*] and Jörn Bonse[2,*]**

[1] Lenzner Research LLC, 125 E Canyon View Dr, Tucson, AZ 85704, U.S.A.

[2] Bundesanstalt für Materialforschung und -prüfung (BAM), Unter den Eichen 87, 12205 Berlin, Germany

e-mails: matthias@lenzner.us ; joern.bonse@bam.de

*corresponding author

## Abstract

In the last decades, the subject of laser-induced damage (LID) moved from a topic of rather scientific interest to a field with sweeping technological applications, then called laser material processing. When using pulsed lasers, the cumulative effect known as incubation is arguably one of the most fundamental features of the processing event. Incubation manifests itself by the fact that the critical fluence $\phi_{th}$ for LID (known as laser-induced damage threshold, LIDT) depends on the number $N$ of pulses exciting one and only one spot on the sample. In most cases, the threshold fluence decreases with $N$ starting from the single-shot ablation threshold $\phi_1$ and remains constant at $\phi_\infty$ for a large number of pulses ($N > N_C$). No ablation or damage occurs for any number of pulses, if the fluence is kept below the multiple-pulse threshold $\phi_\infty$. In contrast, examples where the LIDT increases with the number of pulses have been reported. The latter effect is known as laser conditioning and is for instance advantageously used when ramping up the power of high-power laser systems. Incubation has been described for many types of solids; the motivation for these comprehensive efforts is twofold as in the case of the underlying effect of LID. Firstly, one tries to prevent the damage of (mainly transparent) optical materials in the beam path of high-energy or high-peak-power lasers. Secondly, one deliberately uses this damage for the sculpting of components. In this chapter, we want to introduce the reader to the physics that is thought to work in the background, when incubation takes place. We are going to look at the parameters controlling LID and highlight the peculiarity of the parameter "number of pulses". We will give an overview of the experimental work done in a variety of materials. There are several physical and chemical mechanisms proposed that govern incubation, and there are several mathematical models to describe the behavior of threshold fluence and ablation rate in dependence on the number of pulses. In a few cases, the two (physics and mathematics) are even related to each other. Eventually, we will show the implications that incubation has on real-world laser machining.

# 1. Introduction

# 2. Mechanisms of Incubation

## 2.1 Experimental Evidence for Incubation

## 2.2 Physical and Chemical Effects

# 3. Incubation Models and Scaling Laws

## 3.1 Quantification of Incubation

## 3.2 Laser-Induced Fatigue Damage

## 3.3 Electronic Defect Accumulation

## 3.4 Mid-Gap State Model

## 3.5 Damage upon Heat Accumulation

## 3.6 Generic Model of Incubation

# 4. Implications of Incubation for the Scaling of Laser Processing

## 4.1 Determination of the Lateral Precision

## 4.2 Determination of the Vertical Precision

## 4.3 Material Processing with Bursts of Pulses

# 5. Outlook: Strategies for Handling Incubation Effects

# 1. Introduction

When investigating laser-induced damage (LID) of matter, one finds that material can be ablated by a finite train of identical laser pulses, each of which is not able to cause macroscopic damage on its own. Numerous experimental results point to a cumulative effect throughout the first pulses of the train, during which the material evolves to a state, where it is destroyed by one of the later pulses of equal fluence. In other words: "The final shot in the train of subthreshold pulses causes the catastrophic breakdown in the material destroyed at structural level by precedent irradiation." [Chmel1997]. The rate of accumulation decreases with decreasing laser fluence down to the point where the material is not altered even by an infinite number of incident laser pulses. This 'safe' fluence is an important parameter for optical components like mirrors, lenses, or windows, and is more and more often listed as a crucial parameter in manufacturer's data sheets for these parts. At this 'safe' fluence, repeated irradiation fails to change any physical or chemical property of the material and the lifetime of the irradiated part is limited only by natural aging in any given environment.

This cumulative effect, today known as *incubation*, is arguably one of the most fundamental features of laser processing events. As we will see in the following Sect. 2, it implies several consequences that seem not to be related at first glance.

Incubation manifests itself by the fact that the laser-induced damage threshold (LIDT), for a pulsed irradiation usually given as a fluence (energy per area), depends on the number $N$ of laser pulses exciting one and only one spot on the sample. Experiments, which send a single or $N$ pulses to a single spot on the sample are often called 1-on-1 or $N$-on-1 experiments, respectively. In most cases, the LIDT fluence $\phi_{th}$ decreases with $N$, starting from the single-shot ablation threshold $\phi_1$. It remains constant at $\phi_\infty$ (the 'safe' fluence) for any large number of pulses. Such an accumulative behavior has been found on most materials, including dielectrics [Ashkinadze1966, Wu1981], dielectric thin films [DeShazer1973], semiconductors [BonseAPA2001, Bonse2002Si], metals [Figueira1982, Lee1982], polymers [Aldoshin1979, Sutcliffe1986], and biological materials [Kim2000, Le2016]. Also, the wide-band-gap materials diamond [Kudryashov2005] and sapphire [Stoian2000, Sakurai2019] were found to exhibit cumulative incubation behavior, investigated with the aim to learn about the underlying physics.

However, examples where the LIDT increases with the number of pulses have been reported as well. This effect is known as laser conditioning and may be beneficially used, for instance when ramping up the power of high-power laser systems [Bercegol1999]. The physical processes behind this phenomenon are not necessarily changes ("annealing") inside the irradiated material. They can also include surface cleaning, like redistribution of surface contaminations [WoodBook, p. 194]. Very few cases of materials not exhibiting incubation have been shown, e.g. in bovine cornea [Pettit1991] and polyimide [Pettit1993]. In the following, we will only deal with the case of incubation, where the damage threshold is reduced by irradiation, since it is of practical relevance for material processing applications. The case of "no incubation" will be included in the model calculations by choosing an appropriate value for the parameter(s).

In the historic course of investigation, the effect of incubation was given various names:

- Fatigue [Likhachev1967]
- Accumulation [Eronko1978]
- Pre-threshold phenomenon [Danileiko1976, Tikhomirov1977]
- Precursor to laser damage [Domann1988]
- Cumulative destruction [Khazov1975]
- Irreversible absorption change [Wu1981]
- Structural defect accumulation and relaxation [Banishev2001]

At least the last two denominations demonstrate the attempt to get closer to the physical origins of the phenomenon. The term *incubation* was borrowed from medicine, where it refers to the period between exposure to an infection and the manifestation of symptoms. It is believed that the term was used in 1983 for the first time to describe LID in silicon ("… the length of the incubation period varied inversely with the laser intensity …") [Becker1983]. The concept of incubation is obviously widely used in other fields of physics as well; hence care must be taken not to confuse incubation in LID with other phenomena of the same name, e.g. the incubation phase during the growth of crystalline silicon on amorphous substrates [Chung2011].

Incubation is closely related to the physics of the damage process itself. After the invention and rapid development of the laser in the sixties of the last century, this knowledge entered public scientific awareness slowly but steadily. As often in physics, the course of these investigations is characterized by different interpretations of the observed phenomena, accompanied by changes in the terminology as listed above. In hindsight, most of the observations about multi-pulse LID can consistently be explained by what we now know about incubation. In the following, we will not attempt to provide an exhaustive review of all work done on the subject of incubation (this would be practically impossible), but we will try to present the major paths of experimental and theoretical research.

Early observations noticed a modification of physical properties (like transmission) of optical materials under ongoing irradiation with a high-power laser, without explicitly recognizing this modification as a precursor of optical damage. One example is the consideration of the number of pulses that can be obtained from a laser crystal at a given level of pulse energy, before its degradation. In [Burns1967] for instance, these data were presented for a Q-switched ruby laser generating pulses of 10 ns duration. The "number of shots" is here defined as the number of pulses generated from a ruby crystal before their energy decreases to 30% of the original value. At a level of several GW/cm$^2$ (about 0.1 J/cm$^2$ at 10 ns), the ruby crystal is severely degraded at the first shot. Note that the interaction here is a single-photon process, therefore the relatively low LIDT fluence.

This change of optical properties by repeated laser irradiation was later discovered as being connected to LID, also called "optical breakdown", that occurred as a terminal consequence. As early as 1966, it was stated that *"... the number of exposures necessary to start the breakdown process was inversely related to energy per pulse as compared to the critical energy."* [Ashkinadze1966]. Critical energy here means the single-shot LIDT.

That behavior was investigated and plotted in an $E(N)$ graph as we know it from current publications about the topic. The corresponding graph is shown in Figure 6.1a. The experiments were done in polymethylmethacrylate (PMMA) with 1064-nm pulses of 0.5 ms duration [Likhachev1967]. The observed incubation was labeled "fatigue" and phenomenologically attributed to microscopic cracks generated by sub-LIDT irradiation. Also, it was observed that the lateral extension of the damaged region increases with the number of pulses (Figure 6.1b), a phenomenon that will be explained in the next section. Already in 1967, the wealth of the observed sub-threshold processes caused the notion that the *"... very concepts of 'critical' energy and 'sudden' breakdown must obviously be relative"* [Likhachev1967].

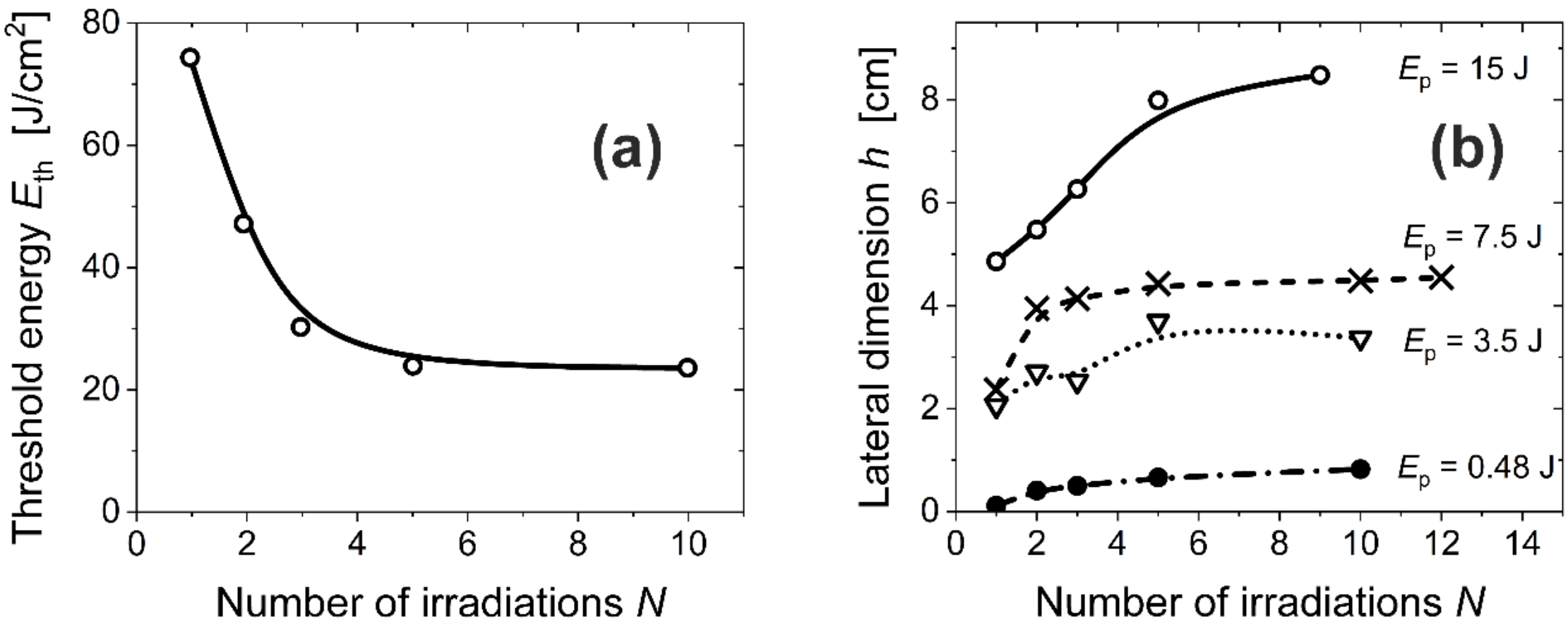

**Figure 6.1: (a)** Dependence of the threshold energy $E_{th}$ on the number of irradiations $N$ in PMMA. **(b)** Dependence of the lateral dimensions of the damaged region $h$ in PMMA on the number of irradiations at various laser pulse energies $E_p$ = 0.48, 3.5, 7.5, and 15 J. Data extracted from [Likhachev1967]. The lines guide the eye

One of the first more systematic studies of the incubation effect was conducted by Eronko et al. [Eronko1978]. Several glass samples were irradiated with a number $N$ of 70-ns pulses, more than an order of magnitude below the single-pulse damage threshold and then investigated by IR reflection spectroscopy. Confirmed by optical microscopy, no damage occurred, but changes in the IR spectrum were observed that were attributed to broken Si-O bonds in the quartz glass. The number of broken bonds increased with the number $N$ of laser pulses but then saturated. Also, the number of broken bonds increased exponentially with the energy density of the applied laser pulses.

Based on these experimental results, Zhurkov et al. gave a concise introduction into the concept of what we call incubation now, taking into account fields of physics other than optics [Zhurkov1980]. They compared the "fatigue" caused by light pulses with (i) the same phenomenon in elasticity (changes in the tensile strength due to repeated mechanical load on a sample) and (ii) the change in AC and DC electrical breakdown strength (AC being of much lower frequency than light here) during extended application of an electric field. The latter effect is known as thermal activation. The authors imported an equation describing this phenomenon from these two subject areas:

$$\tau = \tau_0 \exp\left(\frac{U_0 - \gamma X^*}{kT}\right) \tag{6.1}$$

with $\tau$ – duration of the action until failure and $T$ – temperature. $\tau_0$, $U_0$, and $\gamma$ are material specific parameters. $X^*$ is the physical quantity in question, e.g. the tensile strength $\sigma$, the electric field $E$, or – for LID – the energy density (fluence) $\phi$ or the peak intensity $I_0$ of the applied laser pulses.

Based on this equation, the concept to assign a physical quantity "*damage threshold*" is questioned in [Zhurkov1980]. Quote: *"Rejection of the threshold characteristics of radiation resistance shifts the main attention to the study of irreversible processes preceding macrofracture* [sic]*."* At the time, this has been a rather persistent request, which was obviously not respected during the subsequent research in laser pulse damage. For decades, the laser-induced damage threshold (even being honored with their own acronym LIDT) was reported for a variety of materials and wavelengths without or with only marginal mentioning of the number of pulses applied.

Pointing in the same direction was an approach to consider damage as an inherently probabilistic process at any pulse number and mathematically describe the probability $p_N$ at which $N$ pulses damage a given material in terms of statistics: $p_N = (1 - p_1)^{(N-1)}\, p_1$ [Bass1972]. However, this treatment results in a lower probability for damage at higher pulse numbers which is the opposite of what was later widely observed. Hence this approach was abandoned and the supporting experimental results considered representing extrinsic damage mechanisms like inclusions, cracks, etc. (see e.g. [Jones1989, Danileiko1976]).

Subsequently, even when the probabilistic point of view was not totally abandoned, it turned out that indeed the accumulation of energy by physical or chemical processes in the irradiated material is responsible for an increasing probability for LID at increasing pulse number. A summary of the main suspects for these processes will be given in Sect. 2.2 of this chapter. Manenkov et al. summarized the knowledge about incubation in 1983, holding "... *photochemical, thermochemical, and mechanochemical reactions, and ... different types of phase transitions*" responsible for the accumulation of energy stored in the irradiated material [Manenkov1983]. Based on the available experimental evidence, they eventually established mechanochemical reactions as the main culprit.

From the following time, one finds several publications that noticed incubation taking place, via direct threshold measurements [Merkle1983], via spot-size dependence of the damage threshold [Lee1982, Domann1988], or via decreasing brightness of the laser-induced plasma [Figueira1982]. Tikhomirov and Turovskaya showed signs of irreversible damage visible only under the electron microscope but not leading "... *to appreciable changes in the optical characteristics of the test specimen (transmission, reflection scattering)*" [Tikhomirov1977]. Wu et al., after measuring alterations in alkali halides at 50% of the LIDT suggested that "...*it seems appropriate to redefine the damage threshold as the lowest intensity $I_F$ at which there occurs an irreversible increase in the absorption of materials*" [Wu1981].

Few authors came up with serious physical models like Lee et al. [Lee1983] or Jee et al. [Jee1988], both for metals. The latter paper established a dependence on LIDT and pulse number which was very often used later. It will be reviewed in Sect. 3.2 of this chapter.

In 1997, Chmel [Chmel1997] gave a comprehensive state-of-the-art summary concluding: "*At present there is not a commonly accepted mechanism of the multi-pulse sub-threshold LID. Moreover, the rate of publications on this problem reduced noticeably in recent years although the problem is far from to be resolved* [sic]".

At the end of the last century, when the Titanium:sapphire laser suddenly enabled widespread access to the femtosecond time domain, this situation changed and the number of publications on LID exploded. Around this time, incubation as well returned to attention and became again the subject of scrutiny by the corresponding scientific community, this time with emphasis on the femtosecond range. Soon, the phenomenon of incubation was reported even for the shortest pulses (5 fs) used for ablation until today [Lenzner1999], where the difference between single-shot and 50-shot LIDT fluences turned out to be a factor of 4 for fused silica. Due to the deterministic nature of femtosecond laser damage (i.e. small error bars in the threshold fluence), this factor was far beyond the statistical distribution of the threshold values. Based on more precise experimental data, a number of more sophisticated models for incubation were developed; we will review the most popular of these in Sect. 3 of this chapter.

As a general lesson, it was conceived that when talking about the threshold of damage caused by laser pulses, we should keep in mind that this quantity is a function of the number of applied pulses $N$ and that this function is particularly sensitive (large gradient) if $N$ has just one digit. Also, this effect is more apparent for shorter laser pulses, since for longer pulses, the statistics of the LIDT can easily obscure the effect of pulse-number dependence.

# 2. Mechanisms of Incubation

Laser-induced damage is caused by the absorption of light in the solid. This optically deposited energy is subsequently transferred into other forms of thermal, chemical, or mechanical energy, where it is used for e.g. removal, cracking, changes in stress/strain, etc. Even multiphoton absorption is a continuous process (though having a sharp rise in the dependence of absorption coefficient on light intensity), so the question arises, why there should be a sudden threshold, at which a solid is damaged. For opaque solids, the reason for this lies rather in the physics of the breakdown event within the solid, the absorption mechanism being self-evident. For transparent materials, however, the very mechanism of light absorption itself and the subsequent transformation of the absorbed electromagnetic energy into other forms of energy are paramount objects of dispute. Coarsely, one distinguishes two different LID-causing mechanisms; we will call them “thermal” and “electronic”. Thermal mechanisms can be described with classical thermodynamic equations and occur for damage with cw-lasers, long pulses, or high repetition rates. Often, differential equations for one or more temperatures characterizing the contributing reservoirs are used to model this process and classical thermodynamic criteria, such as melting or evaporation, define a damage threshold. Electronic mechanisms on the other hand, occurring for short and ultrashort pulses, are usually described by semi-classical rate equations and often define a critical electron density as criterion for damage. Generally, it is assumed that for pulses longer than 10 ps, thermal mechanisms govern the ablation process, while for shorter pulses, electronic mechanisms (mainly avalanche and multi-photon ionization) take over and become more prevalent with decreasing pulse duration [WoodBook, p. 62].

In the same way as these underlying "absorption and distribution" mechanisms of LID are discussed and disputed, the mechanisms of incubation are under scrutiny, because one can only investigate the physical action of the second laser pulse on the same site, after it is known in detail, what the first laser pulse did. Naturally, a second laser pulse arriving on a site previously irradiated by a (most often identical) laser pulse is subject to the same physical processes as the first pulse, but in addition sees the changes of the solid that is left behind by the first pulse. These changes can be irreversible or reversible with a certain relaxation time constant, making the LID process dependent on the repetition rate of the laser pulses.

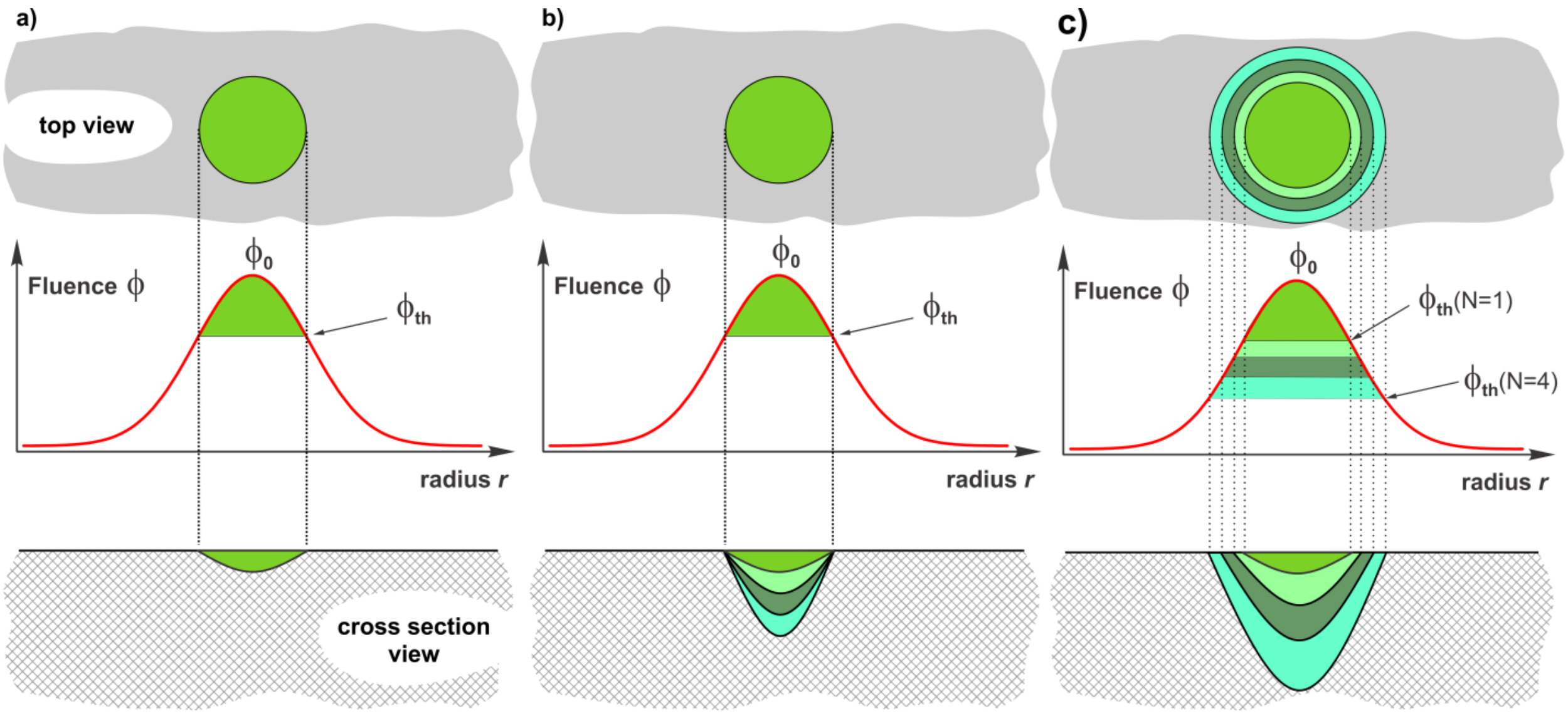

**Figure 6.2:** (**a**) Scheme of surface ablation by a single laser pulse of Gaussian beam profile, top and bottom show the generated crater, the center graph shows the spatial beam profile (fluence vs. radius) of the laser pulse. (**b**) the same in a material that does not exhibit incubation, i.e. the LIDT is constant for all pulse numbers (**c**) the effect of incubation shown for 1, 2, 3, and 4 damaging pulses. Due to the decreasing LIDT, the ablation crater becomes larger in diameter with every laser pulse. The same effect leads to a larger ablation depth than in (b). In all three sub-figures, depth and lateral extension of the crater are not to scale; usually, the ablation depth is much smaller than the lateral size

The principle of LID of a sample surface with a pulsed laser is illustrated in Figure 6.2 by means of three scenarios: (a) single-pulse damage in any material, (b) multiple-pulse damage in a material with a constant LIDT for any pulse number (no incubation), and (c) multiple-pulse damage in a material that exhibits incubation. As mentioned, the second case is a rare one, but it serves well to clarify the principle of incubation. Figure 6.2c points out that the decrease of the LIDT with increasing pulse number results in a lateral effect and an effect in propagation direction. Please note that the depth scale in the cross-section view is strongly exaggerated to better illustrate the effect, the craters are usually much shallower than depicted here.

The qualitative interpretation of Figure 6.2c does not require a mathematically specific dependency of the threshold fluence on the pulse number, as long as it is monotonically decreasing [$\phi_{th}(N_1) > \phi_{th}(N_2)$, for $N_2 > N_1$]. However, the specific dependency does have consequences for the specific mathematical dependency of lateral and depth precision, as we will see below in Sect. 4.

In conclusion, incubation has several implications for the ablation process with multiple pulses, which is why we find several ways to measure and graph these implications.

## 2.1 Experimental Evidence for Incubation

### Implication 1: Size of the Ablation Crater

As shown in Figure 6.2, the diameter as well as the increase in depth of the damage crater depend on the number of pulses. The diameter increase is caused by the fact that a Gaussian beam has a larger diameter at a lower local fluence and the LIDT is decreasing with every additional laser pulse (Figure 6.3a). For larger pulse numbers the damage fluence approaches a constant value and so does the crater diameter. As shown in Figure 6.2b, a constant crater diameter *D* would be the criterion for no incubation. The ablation depth *h* on the other hand increases with increasing excess energy over the threshold, resulting in (i) a non-linear increase of cumulative crater depth with pulse number and (ii) a non-constant incremental crater depth (ablation rate per pulse, Figure 6.3b).

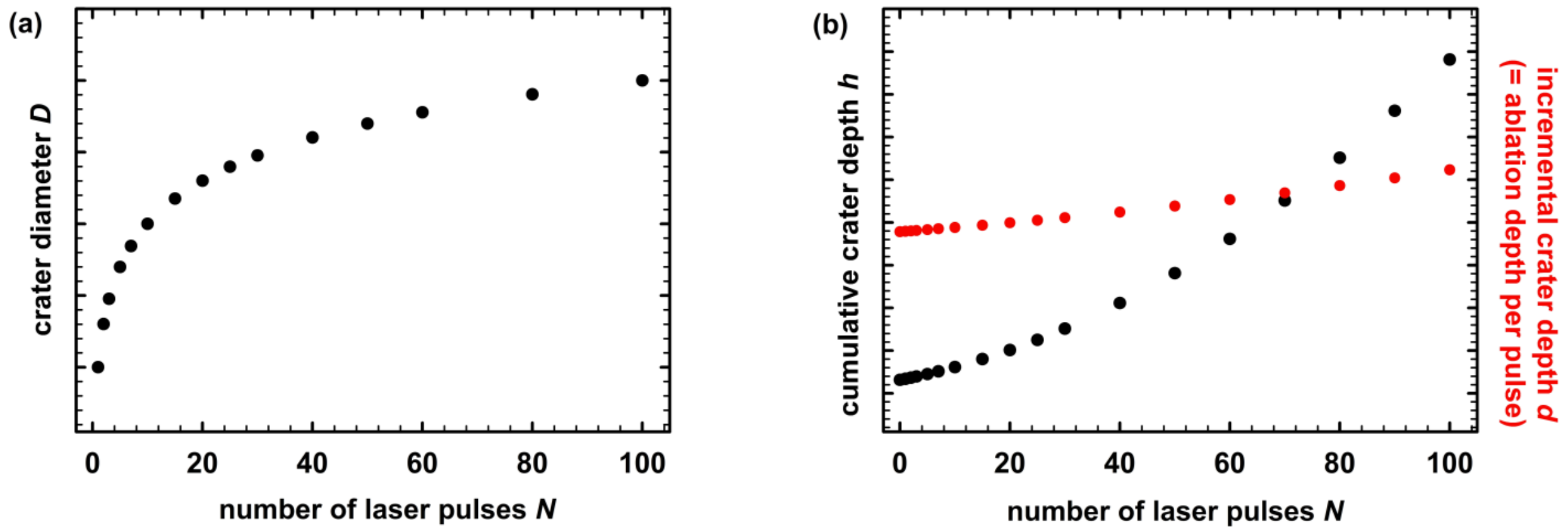

**Figure 6.3: (a)** Non-constant diameter *D* of the ablation crater due to incubation. **(b)** Non-linear increase of the cumulative crater depth *h* (left ordinate) and non-constant ablation depth per pulse *d* (right ordinate) due to incubation

Graphs of this kind have been shown for metals [Mannion2004], semiconductors [BonseAPA2001], Dielectrics [Daengngam2015, Kautek1996], and polymers [Sutcliffe1986, Dorronsorro2008].

For the effect of incubation on the crater depth, it is important to note that this specific non-linear behavior (solely based on incubation) can only be observed for a small number of pulses. The specific value of the "small number" depends on the parameters of the laser pulse (e.g. larger for lower fluence and shorter duration) and of the ablated material (e.g. larger for dielectrics than for metals). For larger pulse numbers, the cumulative crater depth *h* tends to saturate (in other words, the incremental crater depth *d* approaches zero). The effect becomes especially apparent when drilling holes with large aspect ratio (depth much larger than diameter). The reasons for that are manifold:

- If the crater gets deeper, light at the edges of the crater couples at a larger angle of incidence, leading to larger reflection and hence fewer photons in the material.
- The surface tends to get microscopically more irregular with increasing pulse number leading to increased surface scattering and hence fewer photons in the material.
- In a deeper crater, the ablated material cannot be easily ejected and is partly re-deposited at the crater surface, again leading to enhanced surface scattering. This re-deposition is a rather random process that causes some statistical distribution of the crater-depth dependence on the pulse number. It is particularly relevant if the laser processing is conducted in an ambient atmosphere, but less so for processing in a vacuum.

This saturation of the drilling depth has been reported multiple times, e.g. in copper [Neuenschwander2013] (due to decreasing absorption coefficient with $N$), in bovine bone [Cangueiro2012], in Al-Si coatings on steel [Li2015], or in dental tissue [Ji2012, Lukac2016]. In the biological materials, desiccation of the tissue was given as a cause. Since the origins are so diverse and exhibit some statistics, there is no closed theory of this saturation effect. Recently, an empirical function was established as a tool for practical applications [Zemaitis2018].

**Implication 2: Changes in the LIDT**

To explain the crater size effects sketched in Figure 6.2, we presumed that the LIDT is a function of the applied pulse number. In actual fact, this is a corollary of the incubation process and the derivation of this dependency requires extensive measurements. The damage threshold for a series of pulse numbers has to be determined. As we will elaborate later in Sect. 3.1, it is common practice to use Liu's method [Liu1982] to determine the threshold, each of these measurements in turn is a series of crater size measurements extrapolating the crater area / diameter to zero (Figure 6.4a). In spite of this cost, this very embodiment of incubation (Figure 6.4b) is the most common one found in the literature. It is even used to compare LIDT data acquired from identical samples in different laboratories worldwide in round-robin-tests [Starke2003]. Note that this method was developed for single-pulse damage, but it was later shown that it can be used for multiple-pulse ablation as well [Sun2015].

Starting with the single-shot damage threshold $\phi_1$, the threshold fluence decreases with increasing pulse number and finally saturates at $\phi_\infty$. Due to the abundance of experimental data, this curve was the main target of modeling attempts, as we will see in Sect. 3.

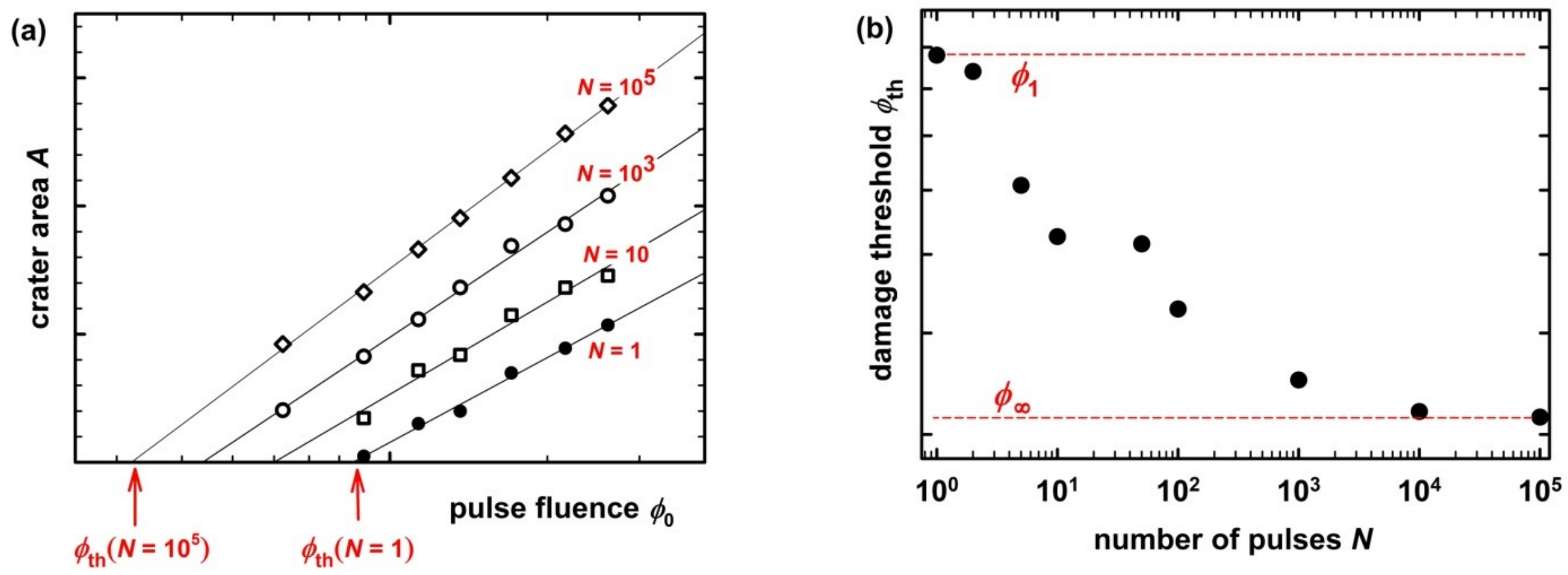

**Figure 6.4:** (**a**) Crater area vs. pulse fluence, used to determine the LIDT for several numbers of pulses. **(b)** The LIDTs from (a), plotted vs. the number of pulses

### Implication 3: Incubation Pulses

An earlier description of incubation stated that, at a given pulse energy, a number of laser pulses $N_C$ (for critical number) has to be applied to prepare (incubate) the material for the upcoming damage at pulse number $N_C+1$. This number of pulses (termed "incubation pulses") was observed to decrease with increasing pulse fluence until it reaches zero at the single-shot LIDT. During the incubation pulses, no damage was observed. However, "observing damage" in many of these cases was a subjective and hence contentious criterion [O'Connell1984, Jones1989]. Later, the pulse-number dependent threshold fluence (Figure 6.4b) became the representation of choice.

The procedure involves an extrapolation as well; here the cumulative crater depth vs. the pulse number obtained from post-irradiation characterization is extrapolated to zero (Figure 6.5a). From this, the fluence dependence of $N_C$ can be determined (Figure 6.5b).

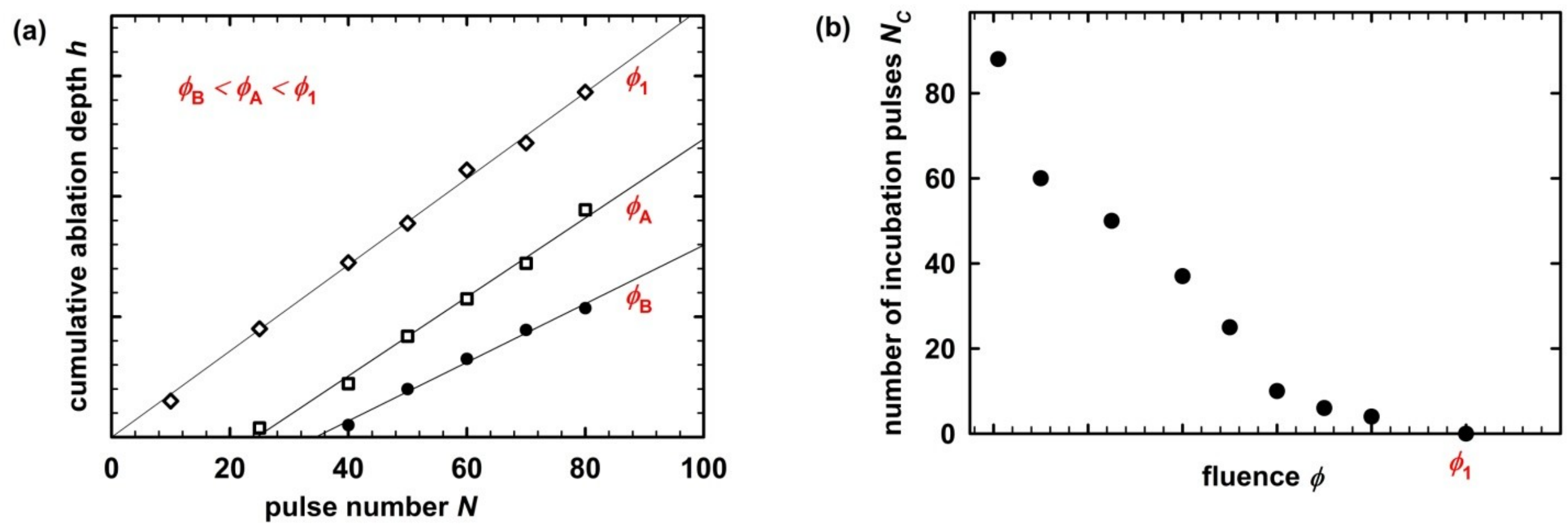

**Figure 6.5:** (**a**) Cumulative ablation depth vs. number of applied pulses for pulse fluences below and at the single-shot threshold fluence. (**b**) Number of incubation pulses in dependence on the pulse fluence, derived as zero crossings from (a). No incubation pulses are necessary at and above the single-pulse threshold fluence $\phi_1$

Jhee et al. [Jhee1985] used *in-situ* electron emission (obviously above the detection limit) on the abscissa, resulting in a critical number of pulses to cause electron emission. Bonse et al.

employed real-time resolved reflectivity measurements during the fs-laser irradiation of polymer films (at fluences moderately below the single-pulse ablation threshold) for quantifying $N_C$ [Bonse2007].

**Implication 4: Effects of the Pulse Repetition Rate**

A consequence of incubation that turned out to be particularly useful for the optimization of material processing is the dependence of LIDT and ablation rate on the repetition rate of the ablating laser pulses. Here, one has to distinguish between irreversible and reversible incubation processes. Only the latter ones will cause an effect when the repetition rate is changed, especially if their relaxation time is in the order of the pulse separation (Figure 6.6a). Faster repetition of the pulses leaves less time for reversible incubation processes to equilibrate (e.g. heat energy to dissipate), hence lower fluences are necessary for damage [DiNiso2014, Raciukaitis2008].

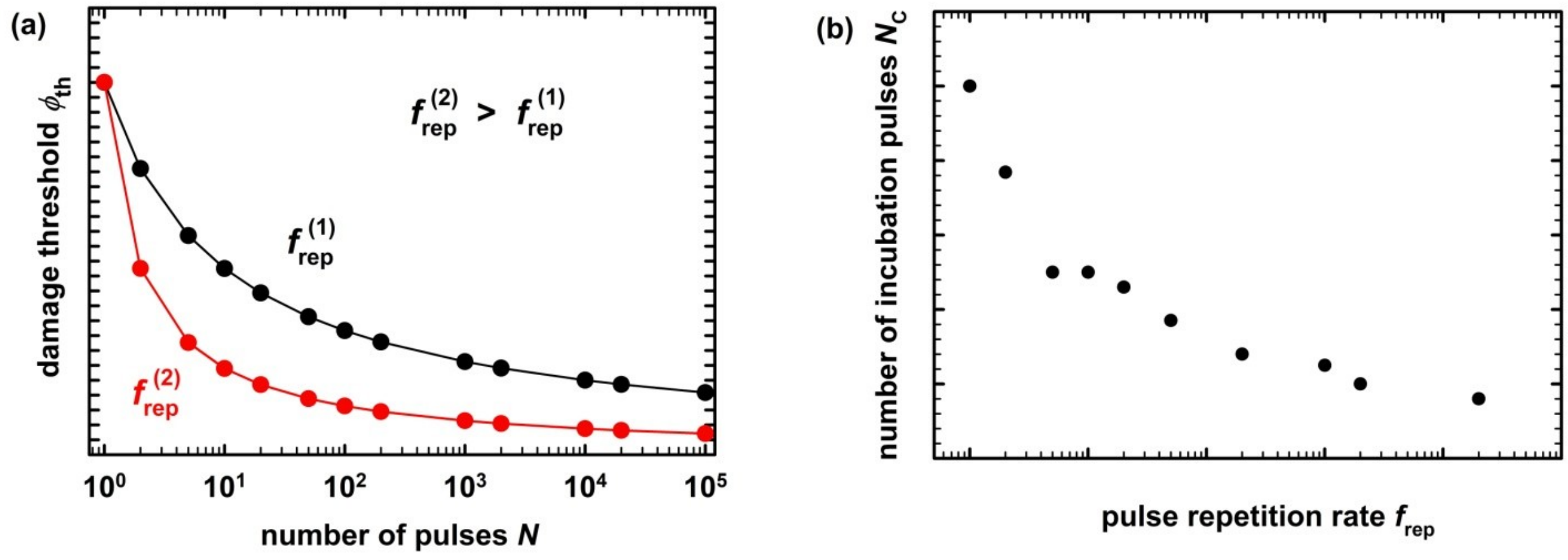

**Figure 6.6:** LIDT vs number of pulses: (**a**) as in Figure 6.4b, but with the pulse repetition rate as parameter. (**b**) Number of incubation pulses in dependence on the pulse repetition rate

In a slightly different approach, Banishev et al. report a dependence of $N_C$ on the pulse repetition rate [Banishev2001], reporting lower $N_C$ for higher repetition rate, which is in accordance with stronger incubation by reversible effects (Figure 6.6b).

**Implication 5: Intra-Pulse Incubation**

For longer pulses (starting from nanoseconds), the below-threshold change of material properties during the pulse can already impact the interaction of the later parts of the pulse with the material. Assuming that the refractive index is one of the material parameters changed during incubation, the simplest method to look at this change is to detect the reflected part of the incident laser pulse with high time resolution (time-resolved reflectometry). The result is that the reflected pulse is "truncated" (i.e. shortened since the trailing edge experiences lower reflectivity) when the laser pulse fluence is above a certain threshold (Figure 6.7). Also, it becomes shorter with increasing fluence in such a way that the reflected fluence (not the intensity!) saturates [Singleton1990]. In PMMA (a material that shows strong incubation), time-resolved reflectometry was used to show a superposition of intra-pulse and pulse-to-pulse incubation. At the same time, this method demonstrated the absence of incubation in polyimide [Pettit1991]. Optical interferometry with higher time-resolution (10 ns) even showed the expansion and contraction of the PMMA surface during the pump pulse [Masubuchi2001]. Fisher et al. showed the same effect for single pulses on bovine cornea, suggesting it as a real-time feedback mechanism for refractive surgery [Fisher2011].

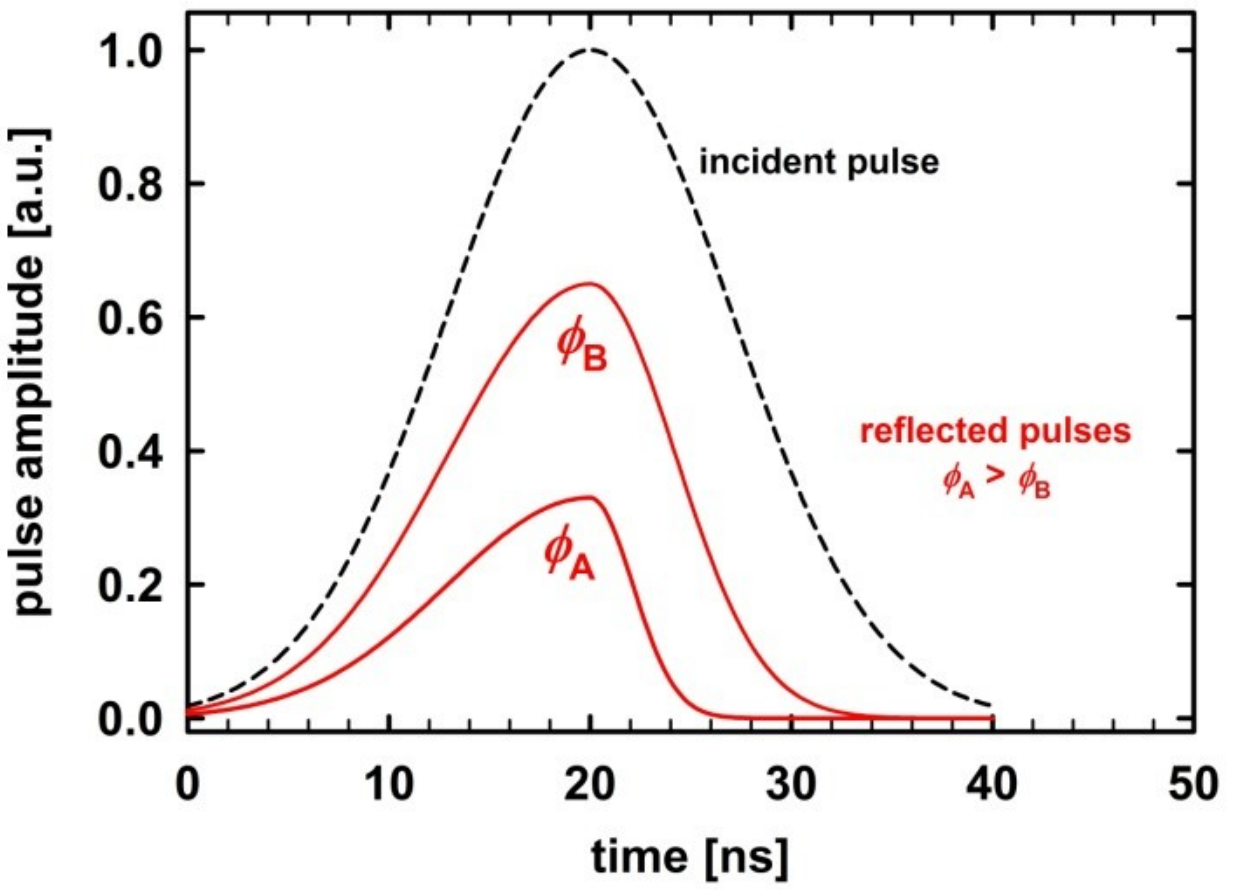

**Figure 6.7:** "Truncation" of the reflected part of an ablating pulse as result of intra-pulse incubation. The ratio of amplitudes of the three curves is not to scale

LID with a continuous wave (cw) laser beam can be considered an asymptotical case of "intra-pulse" incubation with a pulse of infinite duration. Akin to the number of incubation pulses, there is an "incubation period of ablation" (meaning a time period), caused by heating up the sample until there is a "volume explosion" [Bozhevolnyi1993]. As in pulsed incubation, the incubation decreases when the light intensity increases. Using lower power densities, a power-dependent "incubation time" has been observed in lithographic dry development [Oteyza2012]. This knowledge can be used to optimize the speed of laser cutting material, for instance to reduce the laser power necessary to cut through a given thickness of PMMA [Srinivasan1991]. These cw-cases are not to be confused with the "incubation time" for pulsed lasers, often cited instead of the number of incubation pulses at a given fixed repetition rate.

## 2.2 Physical and Chemical Mechanisms

In the quest for the physics/chemistry explaining the accumulation of energy in solids in detail, a smorgasbord of processes was suspected and proven to be responsible. Naturally, different classes of materials exhibit different mechanisms for the absorption and the storage of the incident energy. In this section, we will try to summarize this quest.

### Polymers

The very first observations of incubation in PMMA (Figure 6.1) obviously sparked extensive work about LID in polymers, specifically PMMA (common name: "Plexiglass") and polyimide (common name: "Kapton"). Interestingly, PMMA shows a very strong incubation behavior [Gomez2006], while polyimide has a rather constant ablation rate [Pettit1991].

Aldoshin et al. connect “optical strength” (i.e. resistance to LID) of PMMA to “viscoelastic” properties [Aldoshin1979]. In today’s language that would mean that incubation is a thermal / mechanical process, transporting the energy received by absorption to the molecules neighboring the absorptive inclusion. These in turn store this energy for a time longer than the inverse repetition rate of the laser. Consequently, they can be removed by a smaller energy delivered by the the next arriving laser pulse.

Srinivasan et al. detect the diatomic species $C_2$ and CN to leave the polymer surface with translational energy of about 5 eV, even well below the fluence threshold for significant ns-laser ablation (0.08 J/cm$^2$) at 248 nm [Srinivasan1987]. During the incubation period, the generated pressure might be too small to eject fragments, so that the fragments remain in the ablation crater, mocking the absence of ablation. It was shown that these fragments can be washed out with a solvent.

Küper and Stuke describe UV-laser ablation in PMMA and model the “production of unsaturated species”, preceding the ablation process, by a variable absorption coefficient that depends on pulse-number and fluence [Küper1987, Küper1989]. The model agrees well with experimental results at 16 ns and 300 fs pulse duration. Lippert et al. find an incubation process in which ns-laser pulse absorption at 308 nm increases due to the production of C=C bonds along the polymer backbone [Lippert1999]. Manenkov et al. state that the incubation effect is particularly strong for polymers and showed experiments in PMMA, where thresholds $\phi_1$ and $\phi_\infty$ differ by a factor of 30 [Manenkov1983]. This factum might explain, why intra-pulse incubation – as described in the former sub-section 2.1 (Implication 5) - was until that moment only shown in PMMA.

**Dielectrics**

A significant part of the interest in LID stems from the effort to prevent destruction of optical elements by lasers. Hence, the dielectric substrates of these elements (mirrors, windows, lenses), headed by fused silica and borosilicate glasses, attracted a lot of attention. Two processes lead the list of culprits for intrinsic incubation: (i) the formation of color centers via self-trapped excitons (known from solid-state physics) and (ii) the breaking of bonds between the constituent atoms (known from quantum chemistry). Of course, extrinsic incubation due to the presence of tiny inclusions, pebbles, cracks, or pores in the material are mostly notable in these originally transparent materials, because they can support the generation of conduction-band electrons by single-photon absorption. Another group of publications is dealing with this process, mainly considering the storage of energy in mechanical deformations around these inclusions.

Ashkenasi et al. postulated the formation of color centers that increase absorption at below-bandgap photon energies [Ashkenasi1999]. Color centers are formed either directly from the excited electron-hole pairs or via self-trapped or Frenkel excitons. The group at the University of New Mexico used basically the same model (color center generation via excitons), using the mid-gap states caused by these excitons to solve rate equations for single- and multi-pulse damage thresholds [Emmert2010]. Later, this approach was used to develop a generic model for optical breakdown which will be described in more detail in Sect. 3.6. DePalo et al. found a weak dependence of incubation on pulse repetition rates between 200 kHz down to 60 Hz, arguing that incubation works via defect sites like Frenkel excitons or color centers with a lifetime of about 50 ms [DePalo2022].

Chan et al. used in-situ Raman scattering during irradiation with 130-fs pulses at 800 nm to show large increases in peaks owing to 3- and 4-membered ring structures in the silica network [Chan2001]. This is a very detailed investigation of atomic-scale structural changes in fused silica. Eronko et al. [Eronko1978] blamed broken Si-O bonds for incubation. Danileiko et al. [Danileiko1976] observed pre-threshold increase of nonlinear scattering (i.e. generation of wavelengths different from the irradiating laser), hinting on a refractive index change (and hence structural changes) as incubation mechanism.

Hopper and Uhlmann scrutinized the absorption of light by a metallic inclusion in a glass and what happens next with the absorbed energy [Hopper1970]. They performed detailed calculations about the temperature at the particle-glass boundary, within the particle, and the resulting compression of the glass. Incubation, still called “fatigue” then, was attributed to the melting of the metallic inclusions and their flowing into microscopic cracks. They performed a theoretical permeation of the physics behind LID that has hardly been seen ever since.

Often, precursors to optical damage are investigated with respect to their absorption, scattering, real and imaginary refractive index, and other physical consequences without specifically scrutinizing the physical origin (e.g. bonds or color centers) of the precursor [Gallais2004]. Grigorev et al. performed calculations that map “laser-induced thermostrains” in dependence on the number of pulses and calculate a critical number of pulses for damage from this [Grigorev1992]. Koldunov et al. identified absorbing inclusions as responsible for damage at all pulse durations and hence ascribed incubation to the cumulative collection of energy in these inclusions [Koldunov1998].

Matthias et al. showed one of the few direct observations of an immediate incubation effect by measuring minute changes of absorption in $TiO_2$, $ZrO_2$, and $HfO_2$, when irradiated with a single laser pulse below the LIDT [Matthias1993].

**Semiconductors and Metals**

The nature of energy transfer in laser irradiated metals and semiconductors (in the case of one-photon absorption) is obvious. Therefore, investigations of incubation could concentrate on the subsequent process of energy transport and storage, the latter one leading to incubation.

Especially in metals, thermomechanical explanations of a storage effect are manifest and hence often invoked. For instance, Bernstein et al., when optimizing wire-cutting with laser pulses, found that “*laser-induced thermal expansion initiates mechanical yield above a critical power, $P_{yield}$*” [Bernstein1992]. Even if there were laser-induced changes below that power, they were reversible. At larger powers, mechanical fatigue will lead to damage after a number of pulses, again these processes are considered to be thermal. Banishev et al. identified structural defects in the surface layer (among them vacancies, interstitial sites and dislocations) and their elastic deformation of the lattice as responsible for the accumulation of the energy of below-threshold laser pulses [Banishev2001]. Sato et al. gave the empirical explanation that the LIDT depends on surface roughness and hardness of the material, which is altered by the laser pulses [Sato2013].

Semaltianos et al. investigated LID in a superalloy and stated: "*The incubation effect in metals is known to be due to the formation of a thermally induced plastic stress-strain in the material due to the accumulation of energy from the laser radiation because of its not complete dissipation (during the repetitive pulse irradiation). This stress-strain in turn results in a local plastic deformation of the material providing that the yield strength of the surface is exceeded during each pulse. Such a cumulative effect may also generate surface defects and deteriorate the surface quality of the material which can lead to ablation at a lower threshold.*" [Semaltianos2009]

Lee et al. found thermal stress in copper due to non-uniform heating responsible for the behavior of the multi-pulse damage threshold as well [Lee1982]. In extensive experiments on the spot-size dependence of the LIDT, they connect this heating to thermal diffusion and find that "*... the threshold intensity required to initiate plastic deformation of metal surfaces increases with laser beam size for spots larger than the thermal diffusion length during the pulse*" [Lee1983], giving a useful quantitative measure.

Extensive work on LID in metals was conducted at the University of Texas, Austin. In the beginning, it was stated that the incubation process (then called nucleation) proceeds in a rather heterogeneous way throughout the surface of the material [Sheng1981]. Microscopic explanations couldn't be given at the time. However, later a popular power law (see Sect. 3.2) was established for single crystal metals [Jee1988] and the physics background of this model was explored by considering the equivalent process of mechanical fatigue in Molybdenum [Becker1991]. Naturally, this results in calling thermomechanical processes to account for incubation. The emergence of slip-lines in aluminum and copper and their accumulation under repeated irradiation supported this hypothesis. The energy storage cycle of thermally induced circuits then follows the stress-strain hysteresis loops.

Byskov-Nielsen et al. [Byskov2010] studied the incubation effects in copper, silver, and tungsten by quantifying the threshold fluences, incubation coefficients, and ablation rates upon irradiation by ultrashort Ti:sapphire laser pulses ($\tau_p$ = 100 fs, $\lambda$ = 800 nm). The authors speculated that the surface energy coupling efficiency may increase via a reduction of the surface reflectivity during the first laser pulses. Such topography-mediated effects may additionally open the excitation of resonant optical coupling channels, such as the excitation of *Surface Plasmon Polaritons* (SPPs) or optical near-field enhancement effects [Yang2020].

Wang et al. [Wang2024] presented a coupling model involving hydrodynamic and two-temperature equations for multi-pulse ablation of aluminum and steel, irradiated with femtosecond pulses. "*The ablation on metals by femtosecond lasers is accompanied with* [sic] *complex phase transitions including rapid gasification, matter eruption, metastable superheated liquid, unsteady state and fragmentation.*"

**Miscellaneous Observations**

Cangueiro et al. [Cangueiro2012] found desiccation as responsible mechanism that prepares for ablation of bovine bone. Kim et al. [Kim2000] described incubation in dentin as caused by heat accumulation rather than electronic effects (like buildup of F-centers).

Bonse and Krüger reported that incubation is also visible in the formation of laser-induced periodic surface structures (LIPSS) in silicon, with the fluence range of the occurrence of low-spatial-frequency LIPSS following an incubation curve (fluence vs. number of pulses) parallel to the melting threshold [Bonse2010]. Xu et al. [Xu2020] noted a decreasing LIPSS period with increasing pulse number in a Titanium alloy. These findings are in line with the observations of Balling et al. [Balling2008] since SPPs are often involved in the formation of low-spatial frequency LIPSS on strongly absorbing materials upon irradiation by ultrashort laser pulses [Bonse2009, Bonse2020]. The inter-pulse decrease of the LIPSS period occurs via a grating-assisted SPP-laser coupling mechanism: when the grating-like LIPSS formed during first laser pulses at periods close to the laser wavelength deepens with additional laser pulses, the wavelength of the SPP resonant absorption in the LIPSS grating will undergo a blueshift to smaller SPP wavelength, thus reducing the LIPSS period during multi-pulse exposure [Huang2009].

Zhao et al. [Zhao2023] found a significant effect of incubation in carbon fiber reinforced polymers (CFRP), a currently often used material in lightweight applications with high demands on structural integrity (e.g. airplane fuselages).

# 3. Incubation Models and Scaling Laws

In spite of the differences between different classes of materials, an astonishing similarity of the pulse number dependent LIDT behavior exists, leading to mathematical models that do not dependent on the particularities of the material. Some of these models will be reviewed in the following sub-sections.

## 3.1 Quantification of Incubation

For a quantification of incubation through the measurements of material-specific fluence thresholds, their precise determination and assignment to the specific laser irradiation parameters is essential. There may be very different physical or chemical processes occurring during the laser irradiation, leaving behind characteristic optical or morphological signatures at the sample. Examples of such processes are structural phase transitions like melting, material removal through ablation, oxidation in the ambient air, etc. By carefully inspecting the laser irradiated samples, usually the signatures of these processes can be identified and unambiguously assigned, allowing for the determination of distinct threshold fluences.

Typically, the lowest value in this set of assigned thresholds is referred to as the "laser-induced damage threshold fluence" (LIDT). It represents the lowest fluence where a permanent material modification can be detected by the chosen inspection method (for example optical microscopy). According to an equivalent alternative definition, the LIDT is the highest laser fluence where the damage probability is zero.

Note that the LIDT not necessarily (but often) coincides with the ablation thresholds of the material. Moreover, different sample inspection methods may provide different threshold values as they detect diverse laser-induced changes. For example, optical microscopy relies on sample-induced changes imposed to the inspecting light, while electron microscopy (conducted in a vacuum environment) senses electrons and may be affected by electrical conductivity changes – either laser-induced or due to adsorbed molecules from the environment. Nevertheless, due to its simplicity and wide availability, optical microscopy has become the standard method for characterizing and quantifying laser-induced damage. In the following, we describe the most common methods to quantify damage threshold fluences of laser-irradiated solids for further evaluation of incubation effects.

Concerning the utilized number of pulses for LIDT measurements, several experimental procedures have been established:

- ***1-on-1* irradiation** [Lidaris-1]: A selected site on the sample is irradiated with a single laser pulse at a fixed peak fluence. A fresh sample site is used for each repetition, no matter if the peak fluence is changed or not. A large number of such test sites at varying peak fluence can be used to establish a damage probability for each peak fluence, mitigating statistical fluctuations of the threshold.
- ***N-on-1* irradiation:** A selected site on the sample is irradiated with a specific number $N$ of identical laser pulses. A fresh sample site is used for each repetition, no matter if the peak fluence is changed or not.
- ***S-on-1* test (ISO)** [Lidaris-2]: A selected site on the sample is irradiated with a specific number $S$ of identical laser pulses. A fresh sample site is used for each repetition. During laser irradiation, the sample is monitored for damage (e.g., via optical scattering): if damage is detected after $S_0$ pulses, the irradiation sequence is stopped and $S_0$ is recorded along with the fluence value. This procedure is repeated for a series of peak fluence values to cover the range $0 \leq S_0 \leq S$. Collected damage statistics are then used to calculate the LIDT versus the number of applied laser pulses (exposure dose), according to the industrial standards ISO 21254-1 [ISO-21254-1] and ISO 21254-2 [ISO-21254-2].
- ***R-on-1* (Ramp) test** [Lidaris-3]**:** A specific site on the sample is irradiated with a certain number $R$ of identical laser pulses. If there is no damage, the fluence level is increased and another $R$ pulses are sent to the same site. The test is stopped if damage has been detected at the test site. The corresponding irradiation dose (average peak fluence in the last sequence of $R$ pulses) is recorded. This procedure is only recommended if the number of test sites available at the sample is limited; specifically it does not aim to account for the quantification of incubation effects.

**ISO Standardized Damage Threshold Evaluation**

Given the enormous technological relevance of the subject for the laser industry, an entire industrial and scientific community is studying the laser-induced damage of optical materials and coatings for more than five decades already, as shown by the traditional annual *Boulder Laser Damage Symposium* in the USA. Since the early eighties, systematic *Round-Robin* laser damage test sequences have been organized and reported at the symposium, while the methods for quantifying laser damage and their standardization were actively improved, as reflected by the international standards ISO 11254, developed in the 1990s [ISO-11254-1, ISO-11254-2] and the more recent ISO 21254, dated from 2011 [ISO-21254-1, ISO-21254-2].

The ISO *S-on-1* test represents the current "gold standard" since it enables the most precise quantification of threshold fluences and related incubation effects. However, the required resources (a dedicated test setup with *in-situ* damage detection, measurement protocol, statistical evaluation, etc.) are comparatively high. When selecting $S = 1$, the single-pulse LIDT can be quantified. This single-pulse damage threshold fluence is the material's "intrinsic" LIDT and represents an upper limit of the multi-pulse damage thresholds in case that incubation effects are present. The statistical approach, evaluating damage probabilities for the quantification of damage thresholds, becomes particularly relevant for longer laser pulses, i.e., durations of ns or longer, and for small focal volumes (tight focusing conditions).

In practice, some less-demanding methods are typically used for quantification of incubation.

## Liu's Simple $D^2$-Method

A straightforward (and hence frequently used) method for the determination of the LIDT was proposed by Liu in 1982 [Liu1982]. It relies on the laser treatment of a sample at several test sites, each irradiated with laser pulse of Gaussian beam profile, the energy of which is systematically varied close to and above the damage threshold. A subsequent measurement of the corresponding laser-induced damage spot sizes (e.g. its diameters $D$) yields a graph $D(N)$.

The first step is to plot the squared damage diameters $D^2$ (or equivalently the damaged area) semi-logarithmically as a function of the laser pulse energies $E_p$ of the corresponding damage spot. For a Gaussian beam profile, in this data representation a linear scaling of the data points can be expected, i.e., $D^2 \propto \ln(E_p/E_{th})$. Extrapolation $D^2 \rightarrow 0$ through a least-squares-fit to the data points then yields the damage threshold energy $E_{th}$. Moreover, from the slope of least-squares fit the Gaussian beam spot diameter $2 \cdot w_0$ can be obtained and used in a second step to convert laser pulse energies into peak fluence values ($\phi_0 \propto E_p$, and $\phi_{th} \propto E_{th}$). Given the described procedure, the method is called *"$D^2$-method"*, *"Liu-method"*, or infrequently also *"zero damage method"*.

This procedure can be performed in 1-on-1 (single pulse per spot) or in $N$-on-1 (multiple pulses per spot) modes, yielding single- or multiple-shot LIDTs.

An example for the application of Liu's $D^2$-method is provided and discussed in Figure 1.9 of Chap. 1 (Nolte et al.) for the quantification of the single-pulse ablation threshold of two different optical glasses. The method was successfully applied for the quantification of incubation effects on almost any material including metals, semiconductors, dielectrics, ceramics, polymers, etc. [BonseAPA2001, Bonse2002Si, Wang2024, Zhang2023]. It has become a (non-industrial) standard in the laser processing community.

The $D^2$-method can be easily extended from radially symmetric to elliptical-shaped Gaussian beams, where both principal axes directions can be evaluated separately. In case that the cross-sectional profile of the laser beam used for damage tests is too irregular, a beam-shaping aperture may be placed in front of the beam focusing element, realizing a moderate truncation. The relevance of the resulting far-field diffraction effects on the accuracy of the $D^2$-method was investigated by Garcia-Lechuga and Grojo [Garcia-Lechuga2021].

Apart from being simple, the $D^2$-method has the key advantage of providing self-consistent results, because the Gaussian beam diameter is determined exactly at the sample surface plane. Note that Liu's simple mathematical formalism is strictly valid for single-pulse ablation with a Gaussian pulse, where incubation effects do not have to be considered.

**Extended Liu's Methods – The D-Scan Approach**

Another extension of Liu's method was proposed in 2006 by Samad et al. as *"diagonal-scan"* (*D-scan*) method for the use with ultrashort pulsed lasers [Samad2006]. It is based on the idea to vary the position of the sample surface relative to the beam focus position in axial direction, while simultaneously scanning the beam laterally across the surface. At laser fluences exceeding the material damage threshold, a characteristic bow-tie-shaped damage pattern is left at the surface that can be used to determine (i) the focus position and (ii) the damage threshold fluence of the irradiated material. Oosterbeek et al. extended the D-scan formalism in 2018 to Bessel and vortex beams [Oosterbeek2018].

An important conceptual difference between the $D^2$- and the D-scan methods must be underlined here: in the $D^2$-method, each sample location is always exposed to the same local laser fluence value, in the D-scan method each sample location is exposed to a sequence of different fluence values, depending on the actual Gaussian beam spot size, the selected scanning velocity, and laser pulse repetition frequency. For this reason, Machado et al. employed the concept of locally *"accumulated fluences"* to mathematically approximate incubation effects via the superposition of $N$ irradiation events in such a D-scan [Machado2012]. In parallel, this concept was successfully applied to practical applications of micromachining, deriving the width of a laser-written structure at given beam diameter and sample velocity [Matus2017].

The D-scan method was used to quantify the laser damage or ablation thresholds and incubation effects of numerous materials, such as amalgam and composite resins [Freitas2010], BK7 glass, fused silica, sapphire, [deRossi2012, Machado2012], aluminum [Chang2012], silicon with different doping levels [Oosterbeek2016, Oosterbeek2018], quartz [Oosterbeek2018], GaN [Nolasco2021], and diamond-like carbon [Nolasco2022].

**Comparison of Threshold Determination Methods:**

Table 6.1 compares the three different methods for damage threshold determination and their suitability for quantifying incubation effects, while also considering simplicity, accuracy, and the required technical resources.

| **Criteria / Methods** | **$D^2$-Method** | **D-Scan** | **ISO 21254** |
|---|---|---|---|
| **Simplicity** | ++ | o | -- |
| **Accuracy** | o | - | ++ |
| **Resources** | minimum | medium | maximum |
| **Suitability** for long laser pulses ($\tau_p$ > ns) | o | o | ++ |
| **Suitability** for ultrashort laser pulses (fs - ps) | + | + | ++ |
| **Suitability** for 1-on-1 test | Yes | No | Yes (ISO 21254-1) |
| **Suitability** for *N*-on-1 test | Yes | No | No |
| **Suitability** for *S*-on-1 test | No | No | Yes (ISO 21254-2) |
| **Suitability** for *R*-on-1 test | No | No | No |
| **Axial focus position determination** | No | Yes | No |

**Table 6.1**: Comparison of the three threshold quantification approaches, i.e., the $D^2$-method, the *D*-scan, and the ISO-21254 damage test procedure regarding their suitability for quantifying incubation effects, their simplicity, accuracy, and the required technical resources. Abbreviations: ++: very good; +: good; o: average; -: bad; --: very bad

## 3.2 Laser-Induced Fatigue Damage

Around the early 1980's the experimental observation that optical metal mirrors get damaged upon irradiation with many high-power laser pulses even far below the single-pulse damage threshold has raised the question, whether laser-induced thermomechanical stress degradation (fatigue), caused by cumulative effects of plastic deformation due to thermal stress induced by the repetitive laser heating, may be the origin of this effect [Musal1980]? The technical relevance of this subject has triggered numerous studies that are briefly summarized in this sub-section.

Musal calculated the transient stress-strain response of a plane metal surface under large-spot short-pulse illumination. In that work, the basic equations for surface displacement and fluences for the onset of laser-induced plastic yield were derived [Musal1980]. In 1983, Koumvakalis and co-workers applied Musal's model to multi-pulse laser damage on metal mirrors [KoumvakalisOE1983] and extended their work towards a phenomenological model [Lee1983].

About five years later, Jee and co-workers investigated the damage thresholds and incubation behavior of metal single crystals (copper, aluminum) of different crystal orientation upon laser irradiation with ns pulse durations ($\tau_p$ = 10 ns, $\lambda$ = 1064 nm) [Jee1988]. They proposed a phenomenological model with a simple mathematical expression, describing a reduction of the damage threshold of the irradiated material as a function of the number of laser pulses with constant laser fluence.

The fact that the damage threshold fluence depends on the number of laser pulses, i.e., the material's load accumulates before damage occurs was referred to as *damage accumulation* by the authors instead of *incubation*. The multi-pulse laser damage appeared to be the result of plastic slip deformation accumulating on metal surfaces under repeated irradiation, while following the storage cycle of thermal stress-strain energy induced by the laser pulses. The empirical model of Jee et al. provides a power law that links the single-pulse damage threshold fluence $\phi_{th}(1) = \phi_1$ with the multi-pulse threshold fluence $\phi_{th}(N)$ [Jee1988]:

$$\phi_{\mathrm{th}}(N) = \phi_1 \cdot N^{\xi-1} \tag{6.2}$$

The material-dependent exponent $\xi \in [0,1]$ characterizes the strength of the accumulation. For $\xi = 1$ the threshold does not depend on the number of pulses $N$. The parameter depends on the material as well as on the laser irradiation conditions (wavelength, pulse duration, etc.). Typical values of $\xi$ range from 0.7 to 0.95. For values of $\xi > 1$, even material hardening (laser conditioning) effects can be mathematically described by the model.

For testing their model, the authors conducted *transient photothermal deflection* (TPD) measurements on molybdenum metal mirrors under multi-pulse exposure. Good agreement was found for the ratio $\phi_{th}(N)/\phi_1$ between the prediction of Eq. (6.2) and the experiments with up to $10^4$ laser pulses per site [Becker1991].

The relationship Eq. (6.2) also applies when ultrashort laser pulses are used; there it turned out that it characterizes incubation in the pulse number range $N < 1000$ for a large number of different materials, including metals [Wellershoff1998, Byskov2010, Mannion2004], semiconductors [Bonse2002Si, Raciukaitis2008, BonseAPA2001], ceramics [Bonse2000], glass [Machado2012], dielectrics [Costache2008], polymers [Baudach1999, Baudach2000] . and biological materials [Cangueiro2012, Le2016].

The Jee-model has the key advantage of being mathematically simple. This allows for example to use it to predict a scaling behavior for the laser processing with regard to possible limitations of the lateral and vertical laser machining precision (see Sect. 4 below). An obvious drawback of the model is that the predicted damage threshold fluence vanishes in the limit of a very large number of laser pulse $N$, i.e., $\lim_{N\to\infty}\{\phi_{\mathrm{th}}(N)\} = 0$ for $\xi < 1$. This, however, does not reflect a realistic behavior since in reality a non-vanishing constant saturation value $\phi_{th}(N\rightarrow\infty) = \phi_\infty < \phi_1$ is observed for very large $N$.

This deficiency of the Jee-model was removed by Neuenschwander et al. and Di Niso et al. [Neuenschwander2013, DiNiso2014] by adding the constant offset $\phi_\infty$. Following this idea, we rewrite Eq. (6.2) in the form

$$\phi_{\mathrm{th}}(N) = \phi_\infty + [\phi_1 - \phi_\infty] \cdot N^{\xi-1} \ . \tag{6.3}$$

Simultaneously, Neuenschwander et al. proposed to replace the expression for the *energy deposition depth* $\delta$ (for strong absorbing materials commonly approximated by the optical penetration depth $1/\alpha$) by [Neuenschwander2013]

$$\delta(N) = \delta_0 + \Delta\delta \cdot N^{\psi-1} \, , \tag{6.4}$$

with $\psi$ being another incubation parameter (in analogy to $\xi$).

## 3.3 Electronic Defect Accumulation

The pivotal relevance of material defects, such as cracks, pores, and absorbing inclusions on the laser damage threshold was recognized already very early and discussed for example by Bloembergen in 1973 [Bloembergen1973]. At the same time, de Shazer et al. studied the ns-pulsed ruby laser ($\lambda$ = 694 nm, $\tau_p$ = 8 ns) damage threshold of dielectric thin films and reported a significant dependence of the damage threshold fluence on the laser beam diameter, i.e., their half-wave and quarter-wave thick $ZrO_2$ and ZnS coatings on glass substrates were by a factor of 2 – 3 more resistant to laser radiation when the Gaussian spot radius $w_0$ was reduced from 250 to 52 µm [DeShazer1973], see Figure 6.8. The authors successfully fitted these results with a probability model, assuming that damage relevant microscopic defects or impurities are randomly distributed across the surface and exhibit a lower threshold than the ideal film material. The probability of the laser beam for striking a defect will be greater for larger beam spot sizes, then leading to a reduced damage threshold fluence. It was pointed out that, depending on the nature of the defects, their presence may lead to an increased linear absorption or enhanced multi-photon absorption. For details of the mathematical model, the reader is referred to Ref. [DeShazer1973].

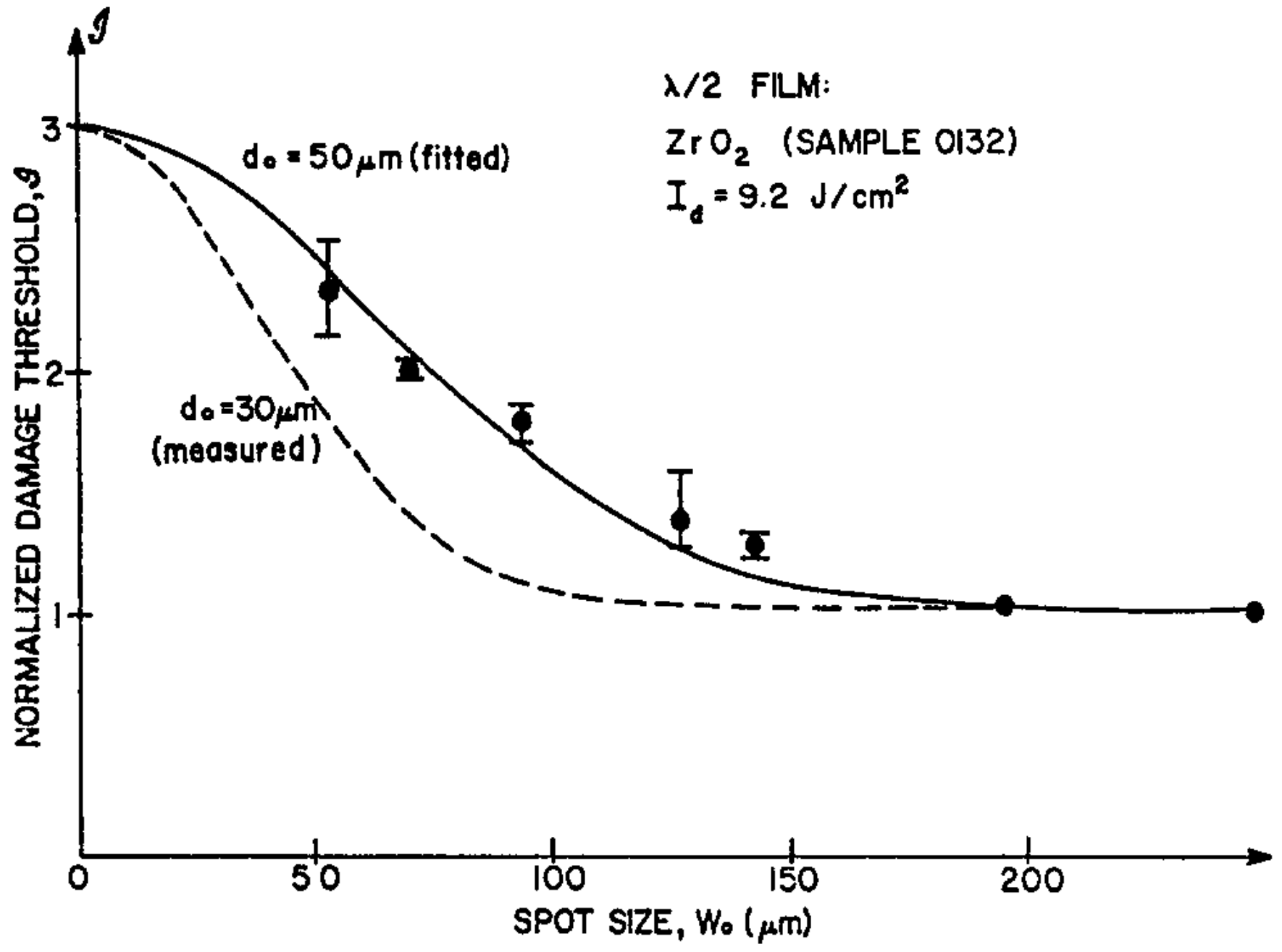

**Figure 6.8:** Spot-size dependence of the damage threshold (normalized) of a half-wave thick $ZrO_2$ coating on glass upon irradiation by single ruby-laser pulses ($\lambda$ = 694 nm, $\tau_p$ = 8 ns) [DeShazer1973]. (Reprinted from [DeShazer1973], DeShazer et al., Role of coating defects in laser-induced damage to dielectric thin films. Appl. Phys. Lett. **23**, 607 (1973), with the permission of AIP Publishing)

In dielectric materials, incubation can also manifest by the pulse-by-pulse accumulation of electronic defect states in band gap of the irradiated material. The generation of such material-specific electronic defects is particularly relevant for irradiation by high intensity laser radiation, as it is readily provided by ultrashort laser pulses. For irradiation of transparent dielectrics with ultrashort laser pulses, the following electronic-defect-mediated incubation scenario can be outlined: already during the high-intensity laser pulse, nonlinear and linear absorption processes, e.g., multi-photon seeded avalanche ionization will result in the development of a free electron plasma in the conduction band of the solid. Recombination of electron-hole pairs or localized trapping will eventually eliminate the electron plasma. Some of the laser-induced electrons may be captured by structural defects or impurities, or they can initiate the formation of meta-stable electronic defect states such as *self-trapped excitons* (STE) in fused silica. With a finite probability such trapped electrons may develop into permanent defects (*color centers*) with energy states inside the band gap. Such additional electronic defect states, in turn, facilitate a lower-order multi-photon excitation by subsequent laser pulses, providing then an enhanced contribution to the electron plasma. The pulse-by-pulse accumulation of electronic defects may then cause the laser-induced electron plasma to reach a critical concentration, leading to surface damage even at lowered laser fluences.

Around the turn of the millennium, another phenomenological incubation model was developed by a group of researchers from the *Max-Born-Institute for Nonlinear Optics and Short Pulse Spectroscopy* in Berlin, Germany, who studied the damage behavior of several wide band gap dielectrics (fused silica, yttrium-lithium-fluoride crystals, sapphire) upon irradiation by single- and multi-pulse ultrashort Ti:sapphire laser pulses with durations between 0.1 and 5 ps [Ashkenasi1999, Ashkenasi2000, StoianPhD2000]. An example of their experimental results obtained for the damage threshold fluence of “gentle” and “strong” ablation of crystalline $Al_2O_3$ (sapphire, corundum) upon variation of the number of laser pulses $N$ for four different laser pulse durations $\tau_p$ between 0.2 ps (a) and 4.5 ps (d) is reprinted in Figure 6.9 [Ashkenasi2000]. Gentle ablation occurs at lower fluence near the ablation threshold and is primarily driven by the rapid creation of an electron plasma within the material, leading to a "Coulomb explosion" where atoms are ejected due to electrostatic repulsion. Strong ablation happens at higher fluence and involves significant heating of the material, causing thermal vaporization and more extensive damage.

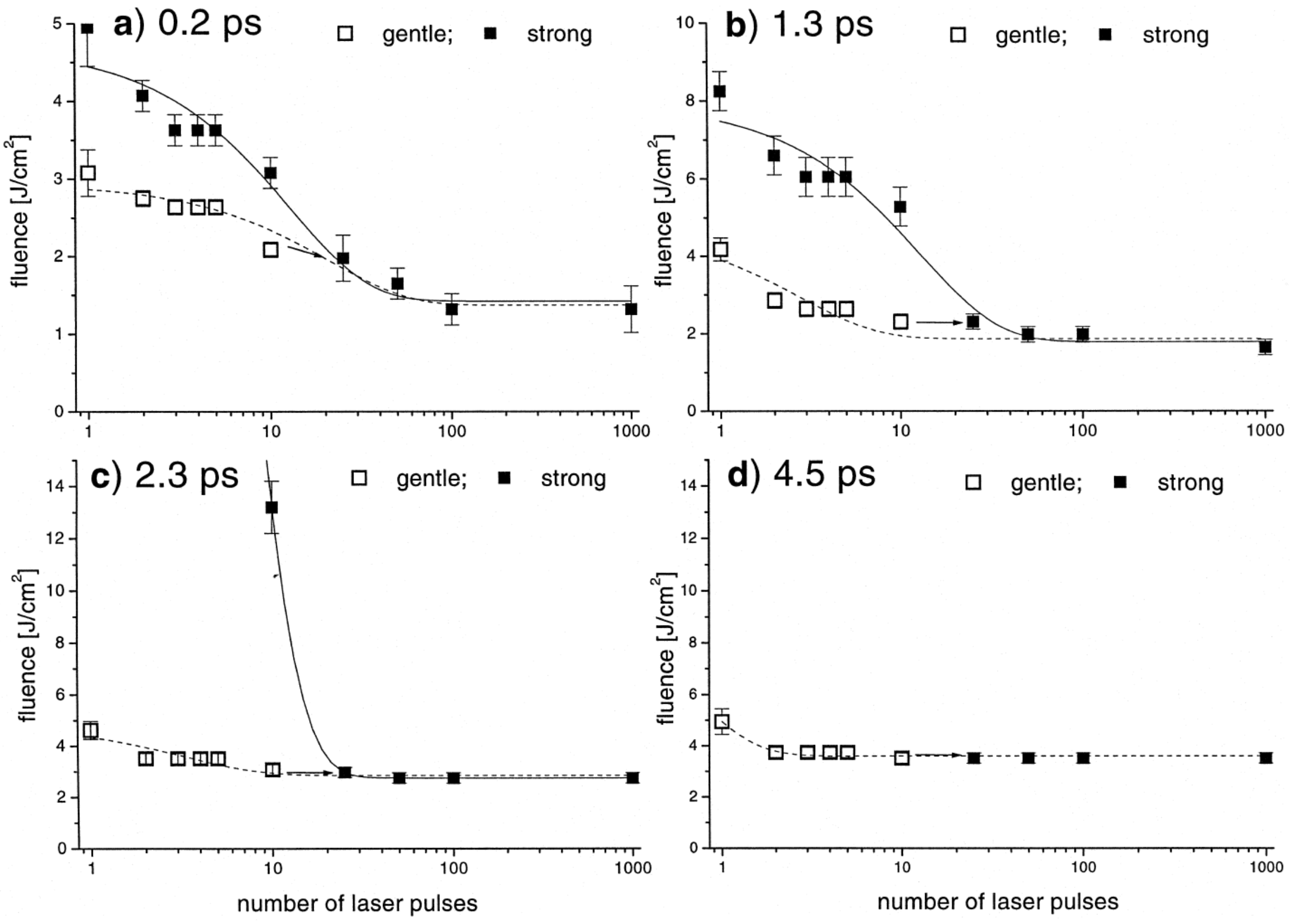

**Figure 6.9:** Semi-logarithmic plot of the surface ablation threshold of the gentle (open squares) and the strong ablation phase (full squares) versus the number of laser pulses per spot for crystalline corundum at four different laser pulse durations $\tau_p$ = 0.2, 1.3, 2.3, and 4.5 ps ($\lambda$ = 800 nm, $f_{rep} \leq 20$ Hz, $10^{-3}$ mbar) [Ashkenasi2000]. (Reprinted from [Ashkenasi2000], Appl. Surf. Sci., **154-154**, D. Ashkenasi, R. Stoian, A. Rosenfeld, Single and multiple ultrashort laser pulse ablation threshold of $Al_2O_3$ (corundum) at different etch phases, 40 – 46, Copyright (2000), with permission from Elsevier)

According to Ashkenasi et al. [Askenasi1999] at a given constant pulse duration, the threshold of surface damage decreases exponentially with an offset that is independent on the number of laser pulses $N$:

$$\phi_{\text{th}}(N) = \phi_\infty + [\phi_1 - \phi_\infty] \cdot \mathrm{e}^{-k(N-1)} \tag{6.5}$$

The exponential coefficient $k$ quantifies the degree of incubation caused by the accumulation of electronic defect states located within the bandgap. This initial hypothesis was further confirmed in Chap. 3 of [StoianPhD2000] by deriving the functional form of Eq. (6.5) using the assumption that the increment in the laser-induced defect concentration upon increasing $N$ is proportional to the product of the initial concentration and the increment in the number of pulses. The laser damage process is then initiated at a certain concentration of active ionization centers. These are required to provide a sufficient number of electrons for exceeding the critical electron density in the conduction band, causing damage of the solid. An alternative derivation of Eq. (6.5) is based on a defect-related rate equation model [StoianPhD2000].

## 3.4 Mid-Gap State Model

Especially for the modeling of femtosecond LID experiments in dielectrics, a simple semiclassical rate equation treatment was often chosen [Du1994, Stuart1996, Lenzner1998]. The electron density in the conduction band $N_e$ was modeled as

$$\frac{d}{dt} N_e(t) = -\alpha_i \cdot I(t) \cdot N_e(t) + \beta_m \cdot I(t)^m. \tag{6.6}$$

where $\alpha_i$ is the impact ionization coefficient, $\beta_m$ is the $m$-photon absorption coefficient, and $I(t)$ is the pulse intensity. Damage was assumed to occur if this electron density exceeds a certain limit (usually $10^{18}$ cm$^{-3}$). It is evident, that this treatment does not take into account any effect of multiple pulses and should therefore only be used to model single-shot data.

Probably the closest approach to the realms of solid-state physics while modeling LID including incubation was undertaken by the group at the University of New Mexico [Mero2005, Emmert2010]. They established a semiclassical model with rate equations for the accumulation and ionization of self-trapped excitons (STEs), where the generation of an STE by a laser pulse supplies an electron that can be excited by the next pulse. In a dielectric, where $m$ photons are necessary to bridge the band gap, this process can be of lower order and can therefore have a higher probability. To this avail, two additional energy states in the forbidden band gap of an insulator were introduced as mid-gap states: shallow and deep traps (see Figure 6.10).

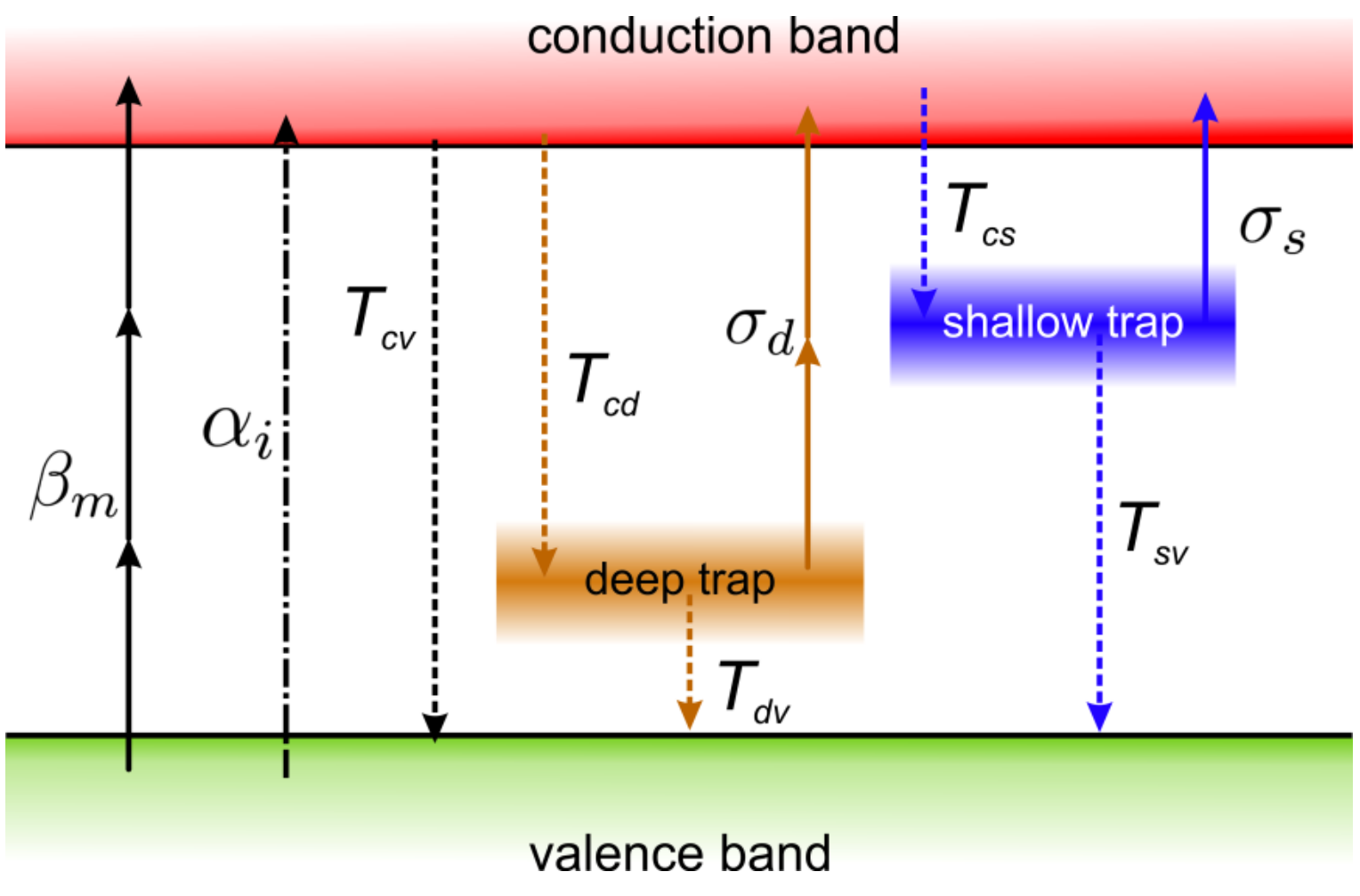

**Figure 6.10:** Schematic diagram of mid-gap energy levels and transitions in a wide-gap insulator relevant for the interaction with a train of laser pulses

Shallow traps are within one photon energy of the conduction band while deep traps are more than one photon energy below. These traps can either be native to the material or laser-induced, then accounting for incubation upon accumulation of such defects. The coefficients in Figure 6.10 are recombination time constants $T_$, absorption cross sections $\sigma_$, and the parameters from Eq. (6.6).

The model now establishes three rate equations for the electron densities $N$, $N_s$, and $N_d$ for conduction band, shallow traps, and deep traps, respectively, taking into account all channels that feed or deplete each level:

$$\frac{d}{dt}N(t) = \alpha_\mathrm{i}\cdot I(t)\cdot N(t) + \beta_m\cdot I(t)^m - \frac{N(t)}{T_\mathrm{cv}}$$
$$-\frac{N(t)}{T_\mathrm{cs}}\left(1-\frac{N_\mathrm{s}(t)}{N_\mathrm{s,max}}\right) - \sigma_\mathrm{s}N_\mathrm{s}(t)\cdot\left(\frac{I(t)}{h\nu}\right)$$
$$-\frac{N(t)}{T_\mathrm{cd}}\left(1-\frac{N_\mathrm{d}(t)}{N_\mathrm{d,max}}\right) - \sigma_\mathrm{d}N_\mathrm{d}(t)\cdot\left(\frac{I(t)}{h\nu}\right)^{m'} \tag{6.7}$$

$$\frac{d}{dt}N_\mathrm{s}(t) = \frac{N(t)}{T_\mathrm{cs}}\left(1-\frac{N_\mathrm{s}(t)}{N_\mathrm{s,max}}\right) - \frac{N_\mathrm{s}(t)}{T_\mathrm{sv}} - \sigma_\mathrm{s}N_\mathrm{s}(t)\cdot\left(\frac{I(t)}{h\nu}\right) \tag{6.8}$$

$$\frac{d}{dt}N_\mathrm{d}(t) = \frac{N(t)}{T_\mathrm{cd}}\left(1-\frac{N_\mathrm{d}(t)}{N_\mathrm{d,max}}\right) - \frac{N_\mathrm{d}(t)}{T_\mathrm{dv}} - \sigma_\mathrm{d}N_\mathrm{d}(t)\cdot\left(\frac{I(t)}{h\nu}\right)^{m'} \tag{6.9}$$

Here, $m$ and $m'$ are the number of photons needed to excite an electron to the conduction band from the valence band and from the deep traps, respectively (in Figure 6.10: $m = 3$, $m' = 2$) and $h\nu$ is the photon energy of the incident laser pulses. As usual, multi-photon excitation channels depend on the intensity of the incident laser pulse, while single-photon channels just depend on the energy. Again, the criterion for breakdown/damage is a critical density of electrons in the conduction band. Interestingly, the specific behavior of the threshold vs. pulse number depends on the traps with the highest concentration, but for longer pulses becomes more sensitive to the density of shallow traps only. The model was successfully used to fit incubation in tantalum oxide films [Mero2005, Emmert2010].

## 3.5 Damage upon Heat Accumulation

An origin of incubation that dominates at cw-laser ablation, long pulses, or high repetition rates is the accumulation of heat in the material. A reduction of the damage threshold can manifest at higher fluences, when the residual heat left in the laser-processed volume does not dissipate between successive laser pulses. Then, depending on the thermal properties of the solid and the specific processing constraints such as pulse repetition frequency and laser spot size, the transient peak temperature in the irradiated spot may rise from pulse to pulse, eventually exceeding the temperatures for melting or even evaporation.

Weber et al. derived a *heat accumulation equation* (Eq. (6.10)) that depends on the dimensionality (1D, 2D or 3D) of the dominant dissipating heat flow (cooling) mechanism into the surrounding [WeberOPEX2014, WeberOPEX2017]. For strongly absorbing materials such as metals, irradiation at moderate fluences, and under loose focusing conditions, 1D heat transport is initially dominant for single-pulse events (see Chap. 1, Nolte et al., Sect. 2.2). For long-lasting multi-pulse exposures, however, 3D cooling can dominate if the above-mentioned thermal accumulation effects start playing a role.

Assuming a certain number of $N_\mathrm{p,tot}$ heating laser pulses modelled for simplicity as a temporal δ-peak at the origin of the heat source $Q_\mathrm{nD}$ and with the laser pulse repetition frequency $f_\mathrm{rep}$, the laser-generated temperature increase $\Delta T_\mathrm{HA,nD}$ due to heat accumulation can be estimated [WeberOPEX2014]. Immediately after the laser processing it accounts to

$$\Delta T_{\mathrm{HA,nD}}(t_{\mathrm{proc}}) = \frac{Q_{\mathrm{nD}}}{\rho \cdot c_{\mathrm{p}} \cdot \sqrt{\left(\frac{4\,\pi\, D_{\mathrm{th}}}{f_{\mathrm{rep}}}\right)^{\mathrm{nD}}}} \cdot \sum_{N=1}^{N_{\mathrm{p,tot}}} \frac{1}{\sqrt{N^{\mathrm{nD}}}}. \qquad (6.10)$$

Herein, "nD" is from the set {1D, 2D, 3D}. The heat source terms $Q_{\mathrm{nD}}$ can be written as $Q_1 = \sigma Q_{\mathrm{heat}}/A$, $Q_2 = \sigma Q_{\mathrm{heat}}/\ell$, or $Q_3 = \sigma Q_{\mathrm{heat}}$, where the parameters $A$ and $\ell$ are the area and the length where the residual laser pulse energy $Q_{\mathrm{heat}}$ is deposited. $\sigma = 1$ when the deposited heat can flow into the complete solid angle, and $\sigma = 2$ when the heat can flow into the half-space only. The residual heat $Q_{\mathrm{heat}}$ left in the irradiated sample per incident laser pulse can be written as [WeberOPEX2017]

$$Q_{\mathrm{heat}} = \eta_{\mathrm{heat}} \cdot \eta_{\mathrm{abs}} \cdot E_{\mathrm{p}} = \frac{\eta_{\mathrm{heat}} \cdot \eta_{\mathrm{abs}} \cdot P_{\mathrm{av}}}{f_{\mathrm{rep}}} \quad . \qquad (6.11)$$

$\eta_{\mathrm{abs}}$ represents the fraction of the incident laser pulse energy ($E_{\mathrm{p}}$) that is absorbed in the sample. The fraction of the residual heat $\eta_{\mathrm{heat}}$ that is not used for ablation depends on material parameters, laser parameters (such as pulse duration, fluence, intensity distribution, and wavelength), and processing parameters (such as structure size, and structure depth). Usually, it must be determined experimentally.

Weber et al. provided approximative solutions of Eq. (6.10) that deviate less than 10% from the analytical expression for $N_{\mathrm{p,tot}} > 3$ and have negligible deviations for $N_{\mathrm{p,tot}} > 100$ [WeberOPEX2017]:

1D: $$\Delta T_{\mathrm{HA,1D}} \approx \frac{Q_1}{\rho \cdot c_{\mathrm{p}} \cdot \sqrt{\frac{4\,\pi\, D_{\mathrm{th}}}{f_{\mathrm{rep}}}}} \cdot \left[2\sqrt{N_{\mathrm{p,tot}}} + C_1\right] \qquad (6.12a)$$

2D: $$\Delta T_{\mathrm{HA,2D}} \approx \frac{Q_2}{\rho \cdot c_{\mathrm{p}} \cdot \frac{4\,\pi\, D_{\mathrm{th}}}{f_{\mathrm{rep}}}} \cdot \left[\ln(N_{\mathrm{p,tot}}) + C_2\right] \qquad (6.12b)$$

3D: $$\Delta T_{\mathrm{HA,3D}} \approx \frac{Q_3}{\rho \cdot c_{\mathrm{p}} \cdot \sqrt{\left(\frac{4\pi D_{\mathrm{th}}}{f_{\mathrm{rep}}}\right)^3}} \cdot \left[\frac{-2}{\sqrt{N_{\mathrm{p,tot}}}} + C_3\right] \qquad (6.12c)$$

The constants $C_{\mathrm{i}}$ were determined numerically and account to $C_1 = -1.46$, $C_2 = 0.58$, and $C_3 = 2.61$, respectively. These approximations (Eq. (6.12)) allow exploring the limits and scaling of heat accumulation effects.

Three typical questions of practical relevance are for the maximum laser power, the maximum pulse number, and the maximum repetition frequency that can be used to avoid heat accumulation effects:

1) *What is the maximum laser power $P_{av,max}$ to stay below a given temperature increase $\Delta T_{limit}$?*

The maximum average incident laser power $P_{\mathrm{av,max}}$ at which a given maximum acceptable temperature increase $\Delta T_{\mathrm{limit}}$ is reached for a given total number $N_{\mathrm{p,tot}}$ of laser pulses, can be approximated by [WeberOPEX2017]

1D: $$P_{\mathrm{av,max,1D}} \approx \frac{C_{\mathrm{mat,1}} \cdot \Delta T_{\mathrm{limit}}}{\sigma} \cdot \frac{A \cdot \sqrt{f_{\mathrm{rep}}}}{2 \cdot \sqrt{N_{\mathrm{p,tot}}} + C_1}, \qquad (6.13a)$$

2D: $$P_{\mathrm{av,max,2D}} \approx \frac{C_{\mathrm{mat,2}} \cdot \Delta T_{\mathrm{limit}}}{\sigma} \cdot \frac{\ell}{\ln(N_{\mathrm{p,tot}}) + C_2}, \text{ and} \qquad (6.13b)$$

3D: $$P_{\mathrm{av,max,3D}} \approx \frac{C_{\mathrm{mat,3}} \cdot \Delta T_{\mathrm{limit}}}{\sigma} \cdot \frac{\sqrt{N_{\mathrm{p,tot}}}}{\sqrt{f_{\mathrm{rep}}} \cdot \left(C_3 \sqrt{N_{\mathrm{p,tot}}} - 2\right)}. \quad (6.13c)$$

The material-specific thermal properties are combined here in the expression

nD: $$C_{\mathrm{mat,nD}} = \frac{\rho \cdot c_{\mathrm{p}} \cdot \sqrt{(4\,\pi\,D_{\mathrm{th}})^{\mathrm{nD}}}}{\eta_{\mathrm{heat}} \cdot \eta_{\mathrm{abs}}} \quad . \quad (6.14)$$

It is seen from Eq. (6.13) that in the 1D-case, the limiting average power increases with the square-root of the pulse repetition frequency. In the 2D case it is independent from the pulse repetition frequency and in the in the 3D-case it is decreasing with its inverse root.

2) *What is the maximum number of laser pulses $N_{p,max}$ to stay below a given temperature increase $\Delta T_{limit}$?*

Equation (6.13) can be used to calculate the maximum number of laser pulses $N_{\mathrm{p,max}}$ that can be applied before a specified heat-accumulative temperature increase $\Delta T_{\mathrm{limit}}$ is reached [Weber OPEX2017]:

1D: $$N_{\mathrm{p,max,1D}} \approx \left(\frac{C_{\mathrm{mat,1}} \cdot \Delta T_{\mathrm{limit}}}{\sigma} \cdot \frac{A \cdot \sqrt{f_{\mathrm{rep}}}}{2\,P_{\mathrm{av}}} - \frac{C_1}{2}\right)^2, \quad (6.15a)$$

2D: $$N_{\mathrm{p,max,2D}} \approx e^{\frac{C_{\mathrm{mat,2}} \cdot \Delta T_{\mathrm{limit}} \cdot \ell}{\sigma\,P_{\mathrm{av}}} - C_2}, \text{ and} \quad (6.15b)$$

3D: $$N_{\mathrm{p,max,3D}} \approx \left(\frac{2\,P_{\mathrm{av}} \cdot \sqrt{f_{\mathrm{rep}}}}{C_3 \cdot P_{\mathrm{av}} \cdot \sqrt{f_{\mathrm{rep}}} - \frac{C_{\mathrm{mat,3}} \cdot \Delta T_{\mathrm{limit}}}{\sigma}}\right)^2. \quad (6.15c)$$

3) *What is the maximum pulse repetition frequency $f_{rep,max}$ to stay below a given temperature increase $\Delta T_{limit}$?*

Usually the laser pulse energy $E_{\mathrm{p}}$ and the number of required laser pulses $N_{\mathrm{p,tot}}$ are both known for the desired laser process. Considering that $P_{\mathrm{av}} = E_{\mathrm{p}} \cdot f_{\mathrm{rep}}$, Eq. (6.15) can then be solved for $f_{\mathrm{rep}}$ for estimating the maximum pulse repetition frequency $f_{\mathrm{rep,max}}$ that can be applied:

1D: $$f_{\mathrm{rep,max,1D}} \approx \left(\frac{E_{\mathrm{p}}}{\vartheta_1(\Delta T_{\mathrm{limit}}, N_{\mathrm{p,tot}})}\right)^2, \quad (6.16a)$$

2D: $$f_{\mathrm{rep,max,2D}} \approx \frac{\vartheta_2(\Delta T_{\mathrm{limit}}, N_{\mathrm{p,tot}})}{E_{\mathrm{p}}}, \text{ and} \quad (6.16b)$$

3D: $$f_{\mathrm{rep,max,3D}} \approx \left(\frac{\vartheta_3(\Delta T_{\mathrm{limit}}, N_{\mathrm{p,tot}})}{E_{\mathrm{p}}}\right)^{\frac{2}{3}}. \quad (6.16c)$$

The substituted functions $\vartheta_{\mathrm{i}}$ depend on the number of required laser pulses $N_{\mathrm{p,tot}}$ and consider the specified heat-accumulative temperature increase $\Delta T_{\mathrm{limit}}$.

## 3.6 Generic Model of Incubation

A generic model for incubation was introduced by Sun et al. [Sun2015], where the term "generic" stands for the attempt to cover a broad range of different materials. For this reason, the model lacks the concrete modeling based on solid-state physics, e.g. relaxation and buildup times of mid-gap states as detailed in [Mero2005] or on thermodynamics as developed in [Hopper1970]. It incorporates two physical mechanisms that change with every successive pulse: absorption $\alpha(N)$ and the critical energy $G(N)$ that has to be introduced to cause ablation. Examples of physical processes leading to an absorption increase during the pulse train are the occupation of mid-gap states, the creation of chromophores, and the formation of surface ripples (LIPSS). Examples of mechanisms decreasing the critical specific damage energy are destabilization of the lattice by laser-induced defects and cracking. Assuming the existence of such critical energy is corroborated by experiments in various glasses [Grehn2014].

It is then assumed that the LIDT is reached when the energy density deposited in the absorption depth reaches a critical value

$$\phi_{\text{th}}(N) = \frac{G(N-1)}{\alpha\,(N-1)}\,. \tag{6.17}$$

For the absorption change during a pulse train (pulse number $N$, pulse fluence $\phi$), the following general ansatz is made:

$$\frac{d}{dN}\alpha(N,\phi) = -\beta(\alpha - \alpha_{max})\cdot\phi\ , \tag{6.18}$$

where $\beta$ is a material parameter describing the efficiency of changing the absorption coefficient toward a maximum $a_{\max}$. This and a similar ansatz for the critical energy $G(N)$ lead to the following solutions:

$$\alpha(N,\phi) = \alpha_0 + \Delta\alpha\left(1 - e^{-\beta\cdot\phi\cdot N}\right) \tag{6.19a}$$

$$G(N,\phi) = G_0 + \Delta G\,\left(1 - e^{-\gamma\cdot\phi\cdot N}\right) \tag{6.19b}$$

with $\alpha_0 = \alpha(N{=}0)$ being the absorption depth of the virgin material and $\Delta\alpha = a_{\max} - a_0$. The specific energy $G$ changes from an initial value $G_0$ to its minimum $G_{\min}$ after a large number of pulses, $\Delta G = G_{\min} - G_0$, and $\gamma$ being the coefficient for the pulse-induced change of the specific energy.

As elaborated in Sect. 2.1, implication 2, the most often experimentally determined incubation graph is the LIDT vs. the number of pulses. From these, the single-pulse threshold $\phi_1 = \phi_{\text{th}}(1) = G_0/\alpha_0$ and the multiple-pulse threshold $\phi_\infty$ can be determined. Using these and the two solutions above, Eq. (6.17) can be written:

$$\phi_{\text{th}}(N,\phi) = \frac{\phi_1 - \left[\phi_1 - \phi_\infty\left(1 + \frac{\Delta\alpha}{\alpha_0}\right)\right]\cdot\left[1 - e^{-\gamma\cdot\phi\cdot(N-1)}\right]}{1 + \frac{\Delta\alpha}{\alpha_0}\left[1 - e^{-\beta\cdot\phi\cdot(N-1)}\right]} \tag{6.20}$$

This expression for the threshold fluence can be used to fit experimental data with the free material parameters $\Delta\alpha/\alpha_0$, $\gamma$, and $\beta$. Details about the mathematical implementation of the model and software to fit experimental data can be found online [LenznerWWW].

Equation (6.20) can also be used to describe a multi-component behavior, i.e. specific energy and absorption changes that comprise a sum of $M$ and $P$ processes:

$$\alpha(N, \phi) = \sum_{k=1}^{P} \alpha_k \left(\beta_k, \Delta\alpha_k, N, \phi\right) \tag{6.21a}$$

$$G(N, \phi) = \sum_{k=1}^{M} G_k \left(\gamma_k, \Delta G_k, N, \phi\right) \tag{6.21b}$$

To a certain extent, this model can explain the principal incubation behavior for a broad class of materials and excitation conditions (pulse duration, wavelength, and repetition frequency), in particular, if multiple incubation components are involved. The authors of [Sun2015] used this model to fit their own experimental data (aluminum, $TiO_2$, and nickel superalloy 625, the latter one shown in Figure 6.11) and the data of other authors in Tungsten [Byskov2010] and stainless steel [DiNiso2014]. Other authors used the model to fit the incubation curves in silicon and copper [Zhang2023] as well as in $As_2S_3$ and $As_2Se_3$ thin films [Paula2024].

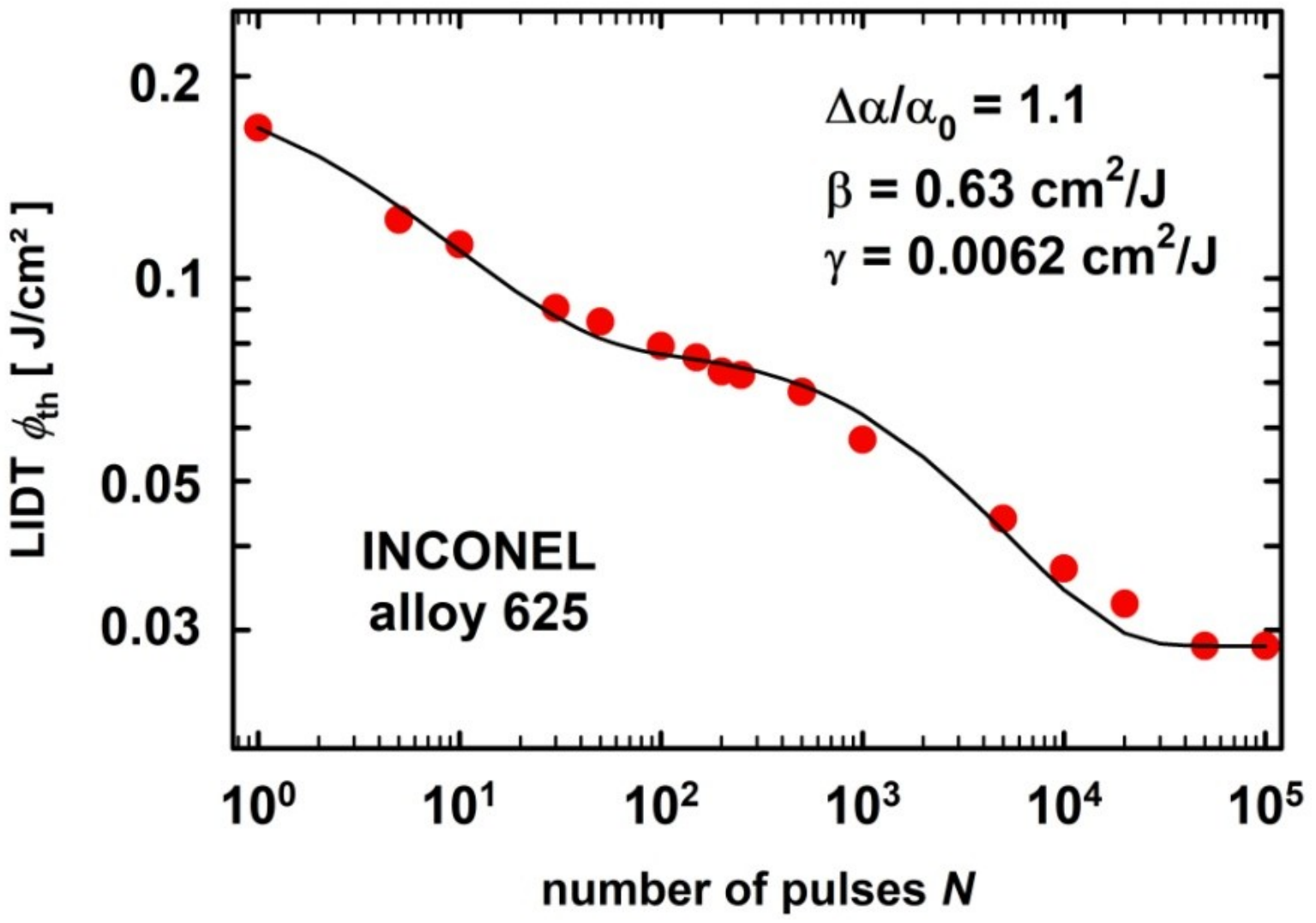

**Figure 6.11:** Incubation curve fitted with the generic model, Eq. (6.20). Red dots: experimental data; solid black line: fitted curve

# 4. Implications of Incubation for the Scaling of Laser Processing

In the following, we discuss the implications of incubation for the laser processing precision, as a summary and extension of Sect. 3.1.4 in [BonsePhD2001]. In this section, the term "damage" can be associated directly with material removal (ablation) but the presented formalism can be generalized also to any other non-ablative material modification that exhibits a sharply defined threshold fluence $\phi_{th}$ (e.g. melt-induced amorphization, recrystallization, oxidation, etc.).

As discussed before, the single-pulse threshold $\phi_1$ is of fundamental physical interest, as it provides insight into the (not yet fully understood) physical processes that occur as a result of laser irradiation, without being obscured by multi-pulse effects. With regard to industrial material processing with laser pulses, however, multi-pulse thresholds $\phi_{th}(N)$ and a precise characterization of the ablation rates become important, since most industrial procedures work with a large number of pulses. Therefore, a compromise has to be found between the irradiated laser pulse energy, the number of pulses per spot, and the pulse repetition rate. Up to here, this chapter summarized the experimental work and the resulting models for the interplay between these physical quantities.

In view of the widely different materials, regardless of the detailed physical processes involved, Eq. (6.2) can be considered an empirical damage accumulation law. It allows to derive simple expressions for the damage spot geometry (diameter and depth), as well as to render conclusions on the lateral (Sect. 4.1) and vertical (Sect. 4.2) laser machining precision obtained with Gaussian laser beam profiles. This may facilitate an assessment of scaling the corresponding laser processing.

## 4.1 Determination of the Lateral Precision

In the model presented here, the lateral extension of a damage site can be described quantitatively by combining Eqns. (1.2) in Chapter 1 and (6.2)

$$D(N, \phi_0) = w_0 \cdot \sqrt{2 \cdot \ln\left(\frac{\phi_0}{\phi_1} \cdot \frac{1}{N^{\xi-1}}\right)} \quad (6.22)$$

The damage diameter $D$ depends on material parameters ($\phi_{th}(1)$, $\xi$) and on laser parameters ($w_0$, $\phi_0$, $N$). To further generalize this equation, we introduce the lateral precision parameter $\chi = D/w_0$ and the normalized fluence $\eta = \phi_0/\phi_1$, yielding

$$\chi = \frac{D(N,\phi_0)}{w_0} = \sqrt{2 \cdot [\ln(\eta) + (1-\xi) \cdot \ln(N)]} \quad (6.23)$$

Two contributions can be identified at the right-hand side of Eq. (6.23), stemming from the laser ($\eta$ and $N$) and the target material ($\xi$), respectively. To illustrate this relationship, Figure 6.12 shows (for typical laser processing conditions) the change in $\chi$ as a function of the number of pulses $N$ for different values of the fluence parameter $\eta$ (a) and for different values of the accumulation parameter $\xi$ (b).

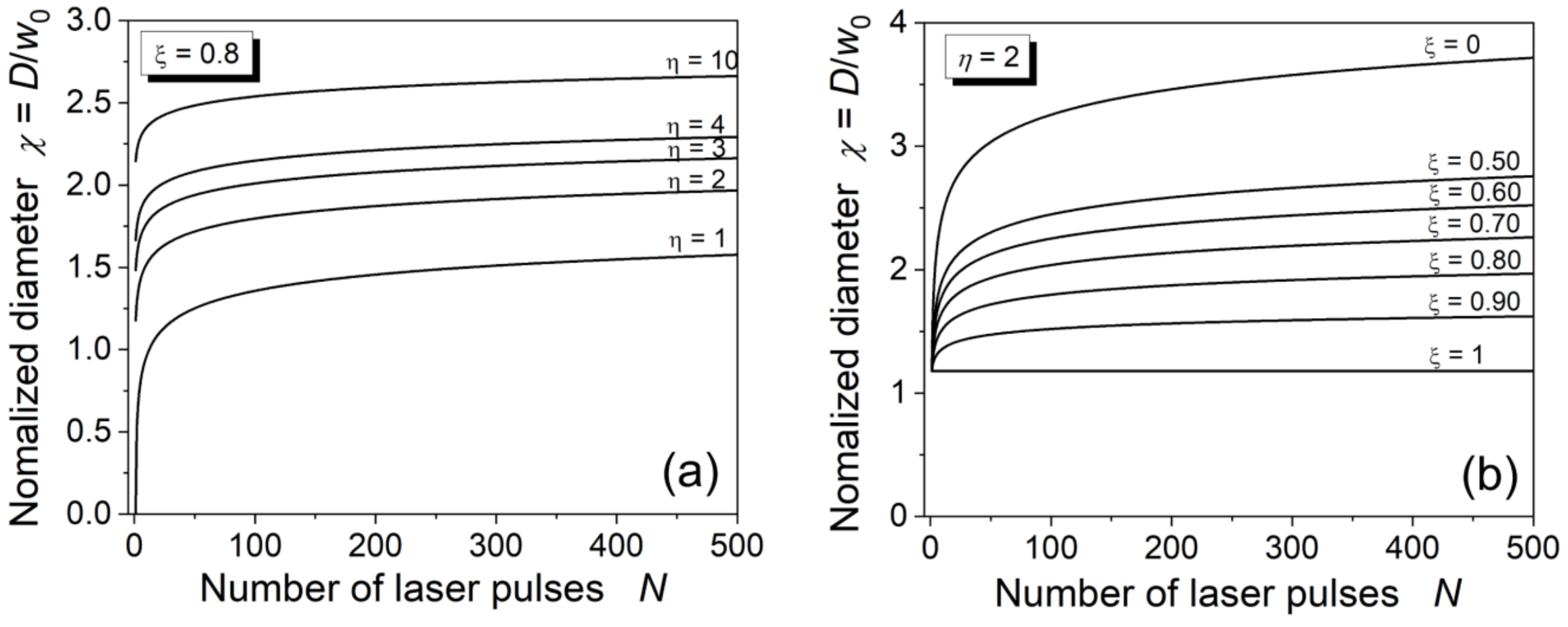

**Figure 6.12:** Variation of the dimensionless damage diameter $c$ according to Eq. (6.23) as function of the fluence ratio $\eta$ (**a**, for a fixed $\xi = 0.8$) and as function as the accumulation parameter $\xi$ (**b**, for a fixed $\eta = 2$)

The dimensionless damage diameter $\chi$ always shows the strongest increase for the first laser pulses, followed by a characteristic ”saturation behavior” for a large number of laser pulses. Figure 6.12a shows the change of $\chi$ with $N$ at constant accumulation parameter $\xi = 0.8$ for different fluence ratio between $\eta = 1$ and 10. For fluences close to the threshold ($\eta < 3$) and for small pulse numbers ($N < 100$), damage diameters smaller than the Gaussian beam diameter ($\chi < 2$) can be achieved. Figure 6.12b shows the variation of $\chi(N)$ at a constant laser fluence of twice the damage threshold for variable accumulation parameters $\xi$ between 0 (strong accumulation) and 1 (no damage accumulation). In the absence of accumulation effects, the damage diameter remains constant, while for $\xi < 1$ a crater widening can always be observed.

We can conclude that incubation counters the generation of arbitrarily small structures by laser pulse processing, regardless of the chosen irradiation fluence.

## 4.2 Determination of the Vertical Precision

As with the lateral precision, incubation has a similar effect on the crater depth and – consequently – the ablation rate. If the crater depth $h$ is considered as a function of the number of pulses $N$ applied per spot, $h$ can be written in the following form for practical use:

$$h(N) = d(N) \cdot (N - N_0) \tag{6.24}$$

This equation defines the pulse number-dependent function $d(N)$, which can be considered the *average ablation rate* per pulse. The quantity $N_0$ denotes the number of *incubation pulses* that are required to achieve material removal. Causes for this offset ($N_0 > 0$) are for instance chemical material modification, surface-adsorbed molecules, altered optical properties, or an increased surface roughness due to the first laser pulses. If $d$ is independent of $N$, each pulse has the same effect on the ablation ($N > N_0$). In general, however, and in the presence of incubation effects, $d$ will depend on $N$.

Considering linear absorption (Beer's law) of a single pulse in the target material, we get the following expression for the *single-pulse ablation depth* $d_1$:

$$d_1(\phi_0) = \delta \cdot \ln\left(\frac{\phi_0}{\phi_{\text{th}}}\right) \tag{6.25}$$

For metals, the proportionality factor $\delta$ should approximately be the value of the light penetration depth $1/\alpha$ in the corresponding material [Preuss1995]. Using Eq. (6.25), the quantity $\delta$ can also be interpreted as the single-pulse ablation rate at a fluence that is $e$ times above the threshold value.

Using the pulse number dependence of the damage threshold fluence $\phi_{th}(N)$ according to Eq. (6.2) and assuming that each pulse ablated the same amount of material, the Eqns. (6.24) and (6.25) can be combined for explicitly obtaining a relationship between the crater depth $h$ and the number of laser pulses per spot $N$:

$$h(N) = \delta \cdot \ln\left(\frac{\phi_0}{\phi_1 \cdot N^{\xi-1}}\right) \cdot N \tag{6.26}$$

However, it must be pointed out that this expression for laser processing in air is only valid for small pulse numbers and crater depths ($h$ < 50 µm), because above a critical crater depth the ablated material is no longer completely ejected from the hole and can partially redeposit on the crater walls (see Implication 1 in Sect. 2.1). In view of the finite extent (diameter) of sufficiently focused radiation, it therefore has to be expected that a saturation behavior in $h(N)$ sets in for sufficiently large crater depths.

With the abbreviations already introduced, a *vertical precision parameter* $\Xi$ (dimensionless crater depth) can be defined as well, measuring the crater depth in multiples of $\delta$:

$$\Xi(N) = \frac{h(N)}{\delta} = [\ln(\eta) + (1-\xi) \cdot \ln(N)] \cdot N \tag{6.27}$$

Figure 6.13 visualizes the scaling behavior of this equation for various fluence ratios $\eta$ (ranging between 1 and 10) for $\xi$ being fixed at 0.8 (a), and for different damage accumulation coefficients $\xi$ (ranging between 0 and 1) for a typical fluence ratio of $\eta = 2$ (b).

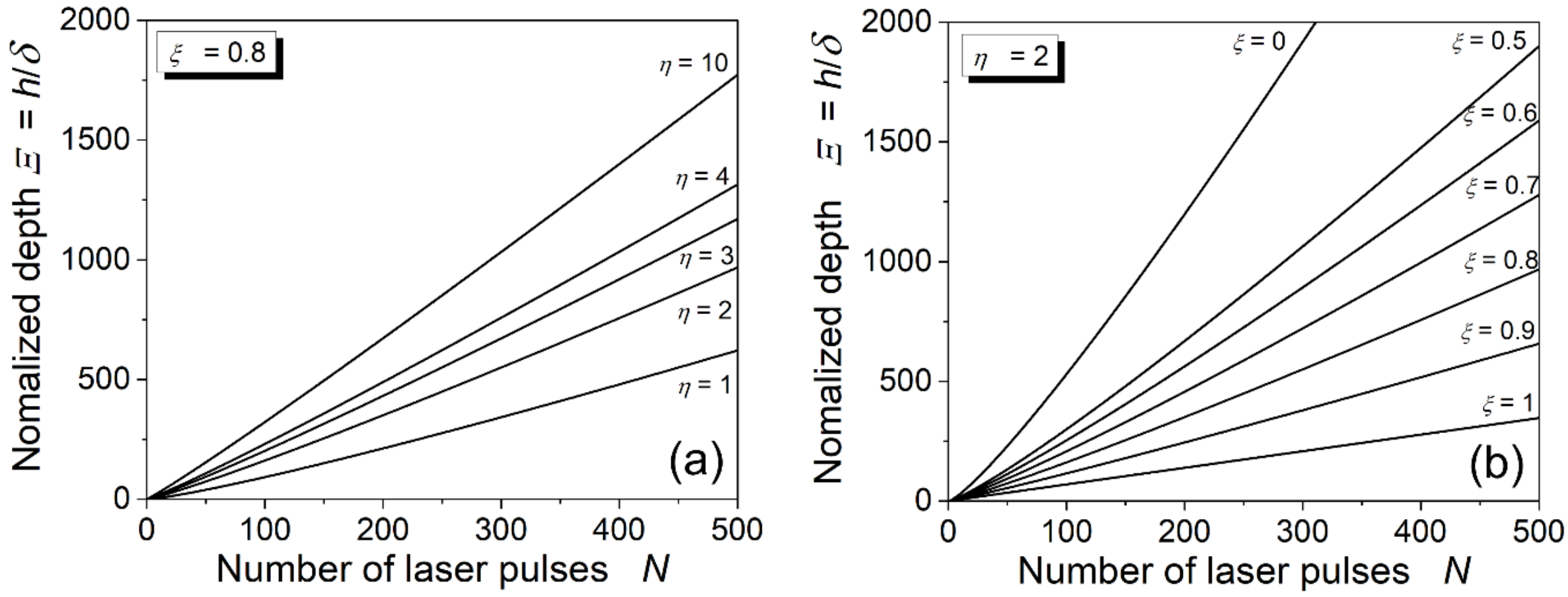

**Figure 6.13:** Variation of the dimensionless crater depth $\Xi$ according to Eq. (6.27) as function of the fluence ratio $\eta$ (**a**, for a fixed $\xi = 0.8$) and as function as the accumulation parameter $\xi$ (**b**, for a fixed $\eta = 2$)

At a fixed damage accumulation coefficient ($\xi < 1$), the normalized crater depth shows a super-linear increase over the number of laser pulses $N$ for all fluence ratios $\eta \geq 1$ (Figure 6.13a). When fixing the laser fluence ratio at $\eta = 2$ – a typical value that enables precise and effective laser processing (see Chap. 5, Neuenschwander et al.) – a linear scaling is observed for $\xi = 1$ only, which is when damage accumulation effects are absent. For all other values $0 \leq \xi < 1$ a super-linear scaling of the normalized crater depths with $N$ is seen as well (Figure 6.13b).

## 4.3 Material Processing with Bursts of Pulses

An interesting since multivalent application of the knowledge about incubation is the processing of target materials with bursts of laser pulses. A small number of laser pulses (the "burst") with high repetition rate (MHz to GHz) is repeated at a lower repetition rate (kHz to MHz). Here, precision and ablation rate can be optimized by separating reversible from irreversible incubation processes.

Early work on this subject was performed around the turn of the millennium by Marjoribanks, Herman, and coworkers at the University of Toronto (Canada), employing pulse train bursts of ultrashort laser pulses for improving the machining precision of dielectrics and metals [Herman1999, Lapczyna1999]. Particularly for dielectric materials, the temporally tailored generation of free carriers in the conduction band is relevant for material processing, because the timescale of subsequent rapid energy relaxation and defect formation processes matches MHz to GHz intra-burst repetition rates. Later, with the development of energetic high repetition rate thin-disk and fiber lasers (see Chap. 2, Gailevicius et al.), the subject of material processing by ultrashort laser pulses with inter-pulse separations in the ns-range gained new interest, adding "ablative cooling" to the realm of hypotheses useful for laser microprocessing [Kerse2016]. Other examples show that efficient GHz rate ablation can be achieved through balancing pulse energies used for heat accumulation incubation to effectively ablate silicon [Mishchik2019] or stainless steel AISI304 [Gaudiuso2018]. Advanced numerical modelling of MHz and GHz burst processing of copper, including incubation effects, was presented by Jia and Zhao [Jia2023].

For strongly absorbing materials, such as metals or semiconductors, Holder et al. developed a comprehensive analytic model for designing efficient process strategies in ultrafast laser micromachining at high average powers and tested it experimentally by employing a state-of-the-art kW-laser system [Holder, 2024]. The authors followed the strategy to maximize the scan velocity and optimize the number of sub-pulses in the burst to avoid heat-accumulation or strong surface roughening. Simultaneously, they equally distributed the high available laser pulse energy over five sub-pulses per burst and dynamically moved the focus in propagation direction during several tens of individual scanning passes. This enabled a high ablation efficiency at sub-pulse peak close to the optimum being $e^2$-times the ablation threshold fluence, see Chap. 5 (Neuenschwander and Förster). By this approach, the team achieved a record ablation rate of $> 2$ mm$^3$/s in silicon, stainless steel, aluminum, and copper

Further sophistication is achieved by combining pulses from different laser sources. Theberge et al. for instance capitalize on incubation to significantly increase the ablated volume of $SiO_2$ by combining pulses from a femtosecond and a nanosecond laser [Theberge2004]. Emmert et al. measured the LIDT for fs-laser irradiation with two pulses of different wavelength (UV and IR) and observed significant differences depending on which of the pulses arrives on the material first [Emmert2015].

Extensive review articles of ultrashort-pulse burst-mode materials processing and the applications in laser surgery are given in [Förster2021] and [Marjoribanks2024]. For high-repetition-rate pulse-burst laser machining, the interested reader is referred to Chap. 16 (Förster et al.) regarding metals and to Chap. 17 (Manek-Hönninger and Lopez) regarding semiconductors and dielectrics.

# 5. Outlook: Strategies for Handling Incubation Effects

Returning to the overall subject of this chapter, i.e. the question how incubation affects laser processing, especially for its up-scaling, the following implications can be summarized:

(i) Understanding incubation is crucial for optimizing laser machining processes and for improving the quality and precision of laser-induced micro-modifications. Here, estimates of the incubation-affected lateral and vertical precision parameters (Eqns. (6.23) and (6.27)), and a detailed analysis of the heat accumulation equation (Eq. (6.10)) may help for scaling the planned laser process towards the desired fluences, repetition rates, and total number of pulses. Usually, a balance between the required precision and energetic process efficiency and the production time and costs must be rendered.

(ii) Incubation affects the durability and lifespan of optical components exposed to repeated laser pulses. Thus, it is essential to consider the extrapolated/estimated LIDT fluence $\phi_\infty$ in the design of lasers and laser-processing machines. This includes also secondary effects from exposure to back-reflected laser radiation, ablation process related short wavelength emissions (UV radiation or even X-rays originating from the ablation plasma), and also the transfer of ablated material (debris). For the latter, the use of protection glasses and tailored gas flows carrying away the ablation material in combination with efficient exhaust systems must be considered and implemented in a scalable manner.

(iii) The detailed interplay between accumulation of electronic defects and accumulation of heat during the incubation phase is widely unexplored yet. Some data for high repetition rates and high laser fluence exist, but mainly from a practical point of view, i.e. to optimize the ablation rate in specific applications without scrutiny of the underlying physics. Especially for dielectrics, it is conceivable that tailoring the repetition rate or applying bursts of pulses can be advantageous, when the repetition periods are fine-tuned with respect to the lifetime of generated electronic defects.

(iv) Similarly, it should be further explored, how the laser-induced ablation plasma created by MHz to GHz rate laser pulse trains could be used to drive molten and evaporated material out of the interaction region, which would lead – among others - to reduced plasma shielding of the incident radiation.

(v) Due to heat accumulation and ablation plasma effects, GHz pulse-burst laser machining often results in significantly smoother surfaces than the ones obtained at lower (kHz to MHz) pulse repetition frequencies, see Chap. 16 (Förster et al.). However, this often comes at the cost of a lower ablation rate. As a remedy, hybrid two-step processing strategies could be considered to optimize the ablation rate in a first efficient material carving step, followed by a second smoothing laser surface finish.

As an outlook, it can be pointed out that detailed research and non-traditional laser processing regimes are currently enabled through the commercial availability of high-repetition-rate ultrafast lasers that additionally permit the generation of tailored pulse bursts. Especially with the advent of multi-GHz repetition rate laser sources, the detailed and knowledge-backed control of incubation processes bears the potential for the acceleration of laser processing to higher production rates.